\pdfoutput=1
\documentclass[prodmode, acmtoms]{acmsmall}
\usepackage[utf8]{inputenc}
\usepackage[T1]{fontenc}
\usepackage{cite}
\usepackage{amssymb}
\usepackage{mathtools}
\usepackage{siunitx}
\usepackage{subcaption}
\usepackage{tikz}
\usepackage{pgfplots}
\usepackage{drawmatrix}
\usepackage{booktabs}
\usepackage{rotating}
\usepackage{pifont}
\usepackage{xspace}
\usepackage{rotating}
\usepackage{aicescover}
\usepackage{hyperref}

\colorlet{graybg}{black!5}
\colorlet{matrixborder}{black!33}
\colorlet{plot1}{red}
\colorlet{plot2}{green!75!black}
\colorlet{plot3}{blue}
\colorlet{plot4}{yellow!75!black}
\colorlet{plot5}{cyan!75!black}
\colorlet{plot6}{violet}
\colorlet{plot7}{orange}
\colorlet{plot8}{brown}

\hypersetup{
    colorlinks=true,       % false: boxed links; true: colored links
    linkcolor=blue!50!black,          % color of internal links
    citecolor=green!50!black,        % color of links to bibliography
    filecolor={.5 0 0},      % color of file links
    urlcolor=red!50!black,           % color of external links
    hypertexnames=false,
}

\def\equationautorefname~#1\null{(#1)\null}

\usetikzlibrary{
    intersections,
    external,
    decorations,
    decorations.pathreplacing
}

\pgfkeys{
    /tikz/external/mode=list and make
}
\tikzsetexternalprefix{figures/externalized/}
\tikzset{external/export=false}

\SendSettingsToPgf
\usepgfplotslibrary{units}
\pgfplotsset{
    compat=1.12,
    every axis/.append style={
        axis background/.style={fill=graybg},
        legend cell align=left,
        xlabel near ticks,
        ylabel near ticks,
        xmin=0,
        ymin=0,
        enlarge x limits={value=0.05, auto},
        enlarge y limits={value=0.05, auto},
        width=\textwidth,
        height=.5\textwidth,
        ymajorgrids=true,
        cycle list name=default,
    },
    twocolplot/.style={
        height=.8\textwidth,
    },
    every axis plot/.append style={
        thick,
        line join=round
    },
}
\pgfplotscreateplotcyclelist{default}{
    {plot1},
    {plot2},
    {plot3},
    {plot4},
    {plot5},
    {plot6},
    {plot7},
    {plot8},
}
\drawmatrixset{
    size=.6,
    fill=graybg,
    draw=matrixborder,
}

\renewcommand\tt\ttfamily
\renewcommand\it\itshape
\renewcommand\sc\scshape

\newcommand\blas{BLAS\xspace}
\newcommand\blasl[1]{BLAS Level~#1\xspace}
\newcommand\lapack{LAPACK\xspace}
\newcommand\relapack{R{\sc e}LAPACK\xspace}

\newcommand\dtrmm{{\tt dtrmm}\xspace}
\newcommand\dtrsm{{\tt dtrsm}\xspace}
\newcommand\dtrtri{{\tt dtrtri}\xspace}
\newcommand\dtrtrir{{\tt dtrtri\_r}\xspace}
\newcommand\dlauum{{\tt dlauum}\xspace}
\newcommand\dpotrf{{\tt dpotrf}\xspace}
\newcommand\dgetrf{{\tt dgetrf}\xspace}
\newcommand\dtrti[1]{{\tt dtrti#1}\xspace}
\newcommand\dpotf[1]{{\tt dpotf#1}\xspace}
\newcommand\xroutine[2][]{{\tt{\it x#1}#2}\xspace}
\newcommand\xlauum{\hyperref[routine:xlauum]{\xroutine{lauum}}\xspace}
\newcommand\xsygst{\hyperref[routine:xsygst]{\xroutine[sy]{gst}}\xspace}
\newcommand\xtrtri{\hyperref[routine:xtrtri]{\xroutine{trtri}}\xspace}
\newcommand\xpotrf{\hyperref[routine:xpotrf]{\xroutine{potrf}}\xspace}
\newcommand\xsytrf{\hyperref[routine:xsytrf]{\xroutine[sy]{trf}}\xspace}
\newcommand\xgetrf{\hyperref[routine:xgetrf]{\xroutine{getrf}}\xspace}
\newcommand\xtrsyl{\hyperref[routine:xtrsyl]{\xroutine{trsyl}}\xspace}
\newcommand\xtgsyl{\hyperref[routine:xtgsyl]{\xroutine{tgsyl}}\xspace}

\newcommand\parsum[1]{\textcolor{orange}{#1}\\}

\renewcommand\parsum[1]{}
\acmVolume{0}
\acmNumber{0}
\acmArticle{0}
\articleSeq{0}
\acmYear{2016}
\acmMonth{2}

\title{
    Recursive Algorithms for Dense Linear Algebra:\\
    The ReLAPACK Collection
}
\aicescovertitle{
    Recursive Algorithms for\\ 
    Dense Linear Algebra:\\
    The ReLAPACK Collection
}
\author{ELMAR PEISE and PAOLO BIENTINESI \affil{AICES, RWTH Aachen}}
\aicescoverauthor{Elmar Peise \and Paolo Bientinesi}
\begin{abstract}
    To exploit both memory locality and the full performance potential of highly
    tuned kernels, dense linear algebra libraries such as \lapack commonly
    implement operations as blocked algorithms.  However, to achieve
    next-to-optimal performance with such algorithms, significant tuning is
    required. On the other hand, recursive algorithms are virtually tuning free,
    and yet attain similar performance.  In this paper, we first analyze and
    compare blocked and recursive algorithms in terms of performance, and then
    introduce \relapack, an open-source library of recursive algorithms to
    seamlessly replace most of \lapack's blocked algorithms.  In many scenarios,
    \relapack clearly outperforms reference \lapack, and even improves upon the
    performance of optimizes libraries.
\end{abstract}


\acmformat{Elmar Peise and Paolo Bientinesi. 2016. 
    Recursive Algorithms for Dense Linear Algebra: The ReLAPACK Collection
}

\begin{document}
    \aicescoverpage

    \begin{bottomstuff}
        Author’s addresses: E.~Peise and P.~Bientinesi, 
        AICES, RWTH Aachen University, Schinkelstr.~2, 52062 Aachen, Germany.
        \{peise,pauldj\}@aices.rwth-aachen.de
    \end{bottomstuff}
    \maketitle

    \section{Introduction}
    \label{sec:intro}
    \parsum{blocked algorithms and their challenges}
Blocking and tiling are common concepts for increasing data locality, reducing
memory stalls, and thus improving performance.  In dense linear algebra, these
concepts are applied to many operations through the means of blocked algorithms
(detailed in \autoref{sec:blocked}) \cite{lapack}, which organize the
computation to attain an extremely favorable ratio of floating point operations
per memory access. Such algorithms offer two degrees of freedom that require
careful tuning: 1) for each operation, there typically exist several algorithmic
variants, which although mathematically equivalent, might differ substantially
in terms of both accuracy and efficiency \cite{trtristability,spdinv}; 2) the
choice of ``block size'' (which ultimately determines how matrices are
traversed) is a critical tuning parameter to achieve nearly optimal
performance~\cite{lapacktuning}.  We stress that the optimal choice for these
degrees of freedom varies (sometimes wildly) with both the computing
environment---the hardware, the implementation of the underlying kernels used as
building blocks, the number of threads used---and the problem size.  As a
consequence, the selection of the algorithmic variant and a quasi-optimal block
size is a tedious and time consuming process.

\parsum{recursive algorithms and \relapack}
For many operations, recursive algorithms\footnote{%
    Algorithms that solve subproblems invoking {\em themselves}; the problem
    size dynamically determines the recursion depth.
} (detailed in \autoref{sec:recursive}) are an alternative that provides
performance comparable to that of blocked algorithms\footnote{%
    Here, blocked algorithms are not considered recursive; they solve
    sub-problems using {\em separate}, unblocked routines.
},
while requiring virtually no tuning effort.  Rather unexpectedly, while blocked
algorithms are readily available in libraries such as the {\sc Linear Algebra
PACKage} (\lapack) \cite{lapack}, hardly any readily available recursive
counterpart exists.  For this reason we introduce the {\sc Recursive \lapack
Collection} (\relapack), an open-source library offering recursive
implementations of many operations.  By conforming to the established
interfaces, these implementations (or a selected subset) can easily replace
\lapack routines in existing software.  Experiments show that \relapack
outperforms \lapack even with optimized block sizes, as well as, in several
scenarios, optimized implementations (such as {\sc MKL} and {\sc OpenBLAS}).

\paragraph{Contributions}  
Our main contribution with this paper is the \relapack library, which provides a
total of 40 recursive algorithms.  It thereby covers almost all of \lapack's
compute routines to which recursion is efficiently applicable; through \lapack's
hierarchical structure, \relapack extends the obtained performance benefits to
over 100 further routines. For all but a few of the operations supported,
\relapack is the first library offering recursive algorithms.  Furthermore, we
provide a detailed analysis of how both blocked and recursive algorithms use
optimized \blas to attain high performance. 

\paragraph{Structure of this paper}
The rest of this paper is organized as follows.  Blocked and recursive
algorithms and the importance of their tuning parameters are illustrated in,
respectively, \hyperref[sec:blocked]{Sections~\ref*{sec:blocked}} and
\ref{sec:recursive}; a performance comparison follows in \autoref{sec:bvsr}.
\hyperref[sec:relapack]{Section~\ref*{sec:relapack}} introduces the \relapack
collection of recursive algorithms, which are compared to high-performance
\lapack implementations in \autoref{sec:vslibs}.  Finally,
\autoref{sec:conclusion} draws conclusions.

    \section{Blocked Algorithms}
    \label{sec:blocked}
    {
\renewcommand\A{\drawmatrix[lower]A\!}
\newcommand\Ainv{\!\!\!\!\drawmatrix[lower]{\hphantom{^{-1}}A^{-1}}}
\newcommand\Azz{\drawmatrix[           fill=plot2!40, lower]{A_{00}}\!}
\newcommand\Aoz{\drawmatrix[height=.4, fill=plot3!40       ]{A_{10}}}
\newcommand\Aoo{\drawmatrix[size=.4,   fill=plot1!40, lower]{A_{11}}\!}

\parsum{blocked algs: idea, variants}
Blocked algorithms extend the performance of the highly optimized {\sc Basic
Linear Algebra Subprograms (\blas)} Level~3 \cite{blas3} kernels to more complex
operations such as matrix inversions, decompositions and reductions.  Each such
operation can generally be implemented as several mathematically equivalent
blocked algorithms, each with a potentially different performance signature.  In
this section, we illustrate how these algorithms operate, and which factors
influence their performance.  We will do so by considering an example: an
algorithm for the in-place inversion of a lower triangular matrix $\A \coloneqq
\Ainv$ in double precision arithmetic.  This operation, known in \lapack as
\dtrtri, and the selected algorithm (there are four alternative variants) are
chosen deliberately simple, but are fully representative of the features and
characteristics of the general class of blocked algorithms.

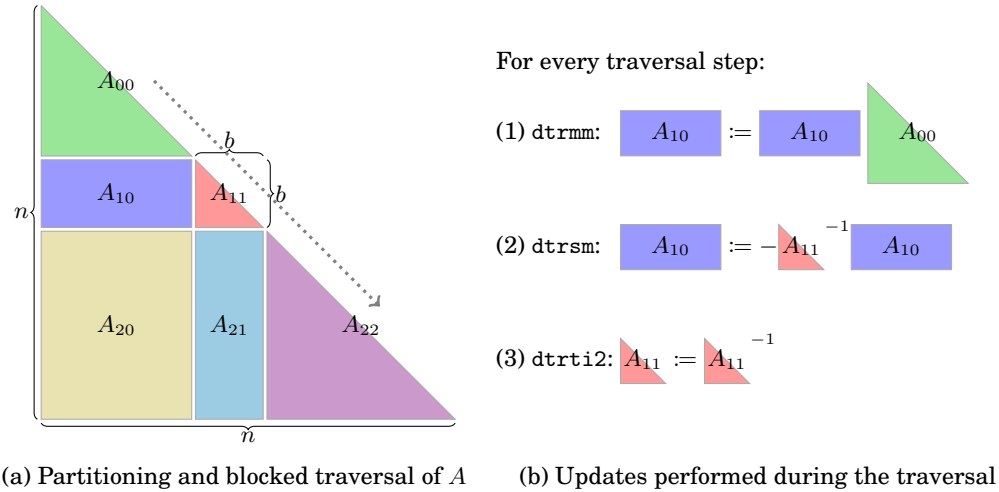
\begin{figure}[t]
    \centering\small

    \renewcommand\Azz{\drawmatrix[height=1.333, width=1.333, fill=plot2!40, lower]{A_{00}}}
    \renewcommand\Aoz{\drawmatrix[height=.6,    width=1.333, fill=plot3!40       ]{A_{10}}}
    \renewcommand\Aoo{\drawmatrix[height=.6,    width=.6,    fill=plot1!40, lower]{A_{11}}}

    \begin{subfigure}[b]{.5\textwidth}
        \centering
        \tikzset{external/export=true}\tikzsetnextfilename{blocked_alg_mat}
        \begin{tikzpicture}
            \filldraw[fill=plot2!40, draw=matrixborder] (0,       -0) --        ++(2,     -2) node[black, midway] {$A_{00}$} -| (0, -0);
            \filldraw[fill=plot3!40, draw=matrixborder] (0,    -2.05) rectangle ++(2,    -.9) node[black, midway] {$A_{10}$};
            \filldraw[fill=plot1!40, draw=matrixborder] (2.05, -2.05) --        ++(.9,   -.9) node[black, midway] {$A_{11}$} -| (2.05, -2.05);
            \filldraw[fill=plot4!40, draw=matrixborder] (0,       -3) rectangle ++(2,   -2.5) node[black, midway] {$A_{20}$};
            \filldraw[fill=plot5!40, draw=matrixborder] (2.05,    -3) rectangle ++(.9,  -2.5) node[black, midway] {$A_{21}$};
            \filldraw[fill=plot6!40, draw=matrixborder] (3,       -3) --        ++(2.5, -2.5) node[black, midway] {$A_{22}$} -| (3, -3);

            \draw[decorate, decoration=brace] (-.05,  -5.5) -- ++(0,    5.5) node[midway, anchor=east]  {$n$};
            \draw[decorate, decoration=brace] (5.5,  -5.55) -- ++(-5.5,   0) node[midway, anchor=north] {$n$};
            \draw[decorate, decoration=brace] (3,    -2.05) -- ++(0,    -.9) node[midway, anchor=west]  {$b$};
            \draw[decorate, decoration=brace] (2.05,    -2) -- ++(.9,     0) node[midway, anchor=south] {$b$};

            \draw[very thick, gray, ->, dotted] (1.5, -1) -- ++(3, -3);
        \end{tikzpicture}
        \tikzset{external/export=false}

        \caption{Partitioning and blocked traversal of $A$}
        \label{fig:blocked_alg:mat}
    \end{subfigure}%
    \begin{subfigure}[b]{.5\textwidth}
        For every traversal step:

        \setlength\tabcolsep{0pt}
        \begin{tabular}{ll}
            (1) \dtrmm:     &\ $\Aoz \coloneqq \Aoz\;\Azz       \vphantom{\drawmatrix[height=1.5]X}$\\
            (2) \dtrsm:     &\ $\Aoz \coloneqq -\Aoo^{-1} \Aoz  \vphantom{\drawmatrix[height=1.5]X}$\\
            (3) \dtrti2:    &\ $\Aoo \coloneqq \Aoo^{-1}        \vphantom{\drawmatrix[height=1.5]X}$\\[10mm]
        \end{tabular}

        \caption{Updates performed during the traversal}
        \label{fig:blocked_alg:ops}
    \end{subfigure}

    \caption{%
        Blocked algorithm for the inversion of a lower triangular matrix.
    }
    \label{fig:blocked_alg}
\end{figure}

\parsum{algorithm 1 in detail}
As shown in \autoref{fig:blocked_alg:mat}, the algorithm traverses the lower
triangular $n \times n$ input matrix $\A$ diagonally from the top left to the
bottom right in steps of a prescribed {\em block size} $b$.  At each step of the
traversal, the algorithm exposes the sub-matrices shown in
\autoref{fig:blocked_alg:mat} and makes progress by applying the three
computational updates in \autoref{fig:blocked_alg:ops}.  Before the execution of
these updates, the sub-matrix $\Azz$ (which in the very first step is of size $0
\times 0$) already contains a portion of the inverse; after the updates, the
algorithm progressed such that the sub-matrices $\Aoz$ and $\Aoo$ now also
contain their parts of the inverse, and in the next step become part of $\Azz$.
Once the traversal reaches the bottom right corner (i.e., $\Azz$ is now of size
$n \times n$), the entire matrix is inverted.

\begin{figure}[t]
    \centering\small

    \ref*{leg:blocked_breakdown}

    \vspace\medskipamount

    \tikzset{external/export=true}
    \begin{subfigure}[t]{.5\textwidth}
        \tikzsetnextfilename{blocked_breakdown_time}
        \begin{tikzpicture}
            \begin{axis}[
                twocolplot,
                ymax=200,
                xlabel={block size $b$},
                ylabel={time},
                y unit=\si\ms,
                stack plots=y,
            ]
                \addplot[plot2, fill, fill opacity=.5] table[y=dtrmm] {figures/data/blocked_breakdown/time.dat} \closedcycle;
                \addplot[plot3, fill, fill opacity=.5] table[y=dtrsm] {figures/data/blocked_breakdown/time.dat} \closedcycle;
                \addplot[plot1, fill, fill opacity=.5] table[y=dtrti2] {figures/data/blocked_breakdown/time.dat} \closedcycle;
            \end{axis}
        \end{tikzpicture}
        \caption{Execution time within the algorithm}
        \label{fig:blocked_breakdown:time}
    \end{subfigure}%
    \begin{subfigure}[t]{.5\textwidth}
        \tikzsetnextfilename{blocked_breakdown_eff}
        \begin{tikzpicture}
            \begin{axis}[
                twocolplot,
                ymax=100,
                xlabel={block size $b$},
                ylabel={efficiency},
                y unit=\si\percent,
                legend to name=leg:blocked_breakdown,
                legend columns=-1,
            ]
                \addplot[plot2] table[y=dtrmm] {figures/data/blocked_breakdown/eff.dat};
                \addlegendentry{\dtrmm}
                \label{plt:blocked_breakdown:dtrmm}
                \addplot[plot3] table[y=dtrsm] {figures/data/blocked_breakdown/eff.dat};
                \addlegendentry{\dtrsm}
                \label{plt:blocked_breakdown:dtrsm}
                \addplot[plot1] table[y=dtrti2] {figures/data/blocked_breakdown/eff.dat};
                \addlegendentry{\dtrti2}
                \label{plt:blocked_breakdown:dtrti2}
            \end{axis}
        \end{tikzpicture}
        \caption{Efficiency within the algorithm}
        \label{fig:blocked_breakdown:eff}
    \end{subfigure}
    \tikzset{external/export=false}

    \caption{%
        Breakdown of the blocked inversion of a lower triangular matrix of size
        $n = 2000$ with increasing block size~$b$.
    }
    \label{fig:blocked_breakdown}
\end{figure}
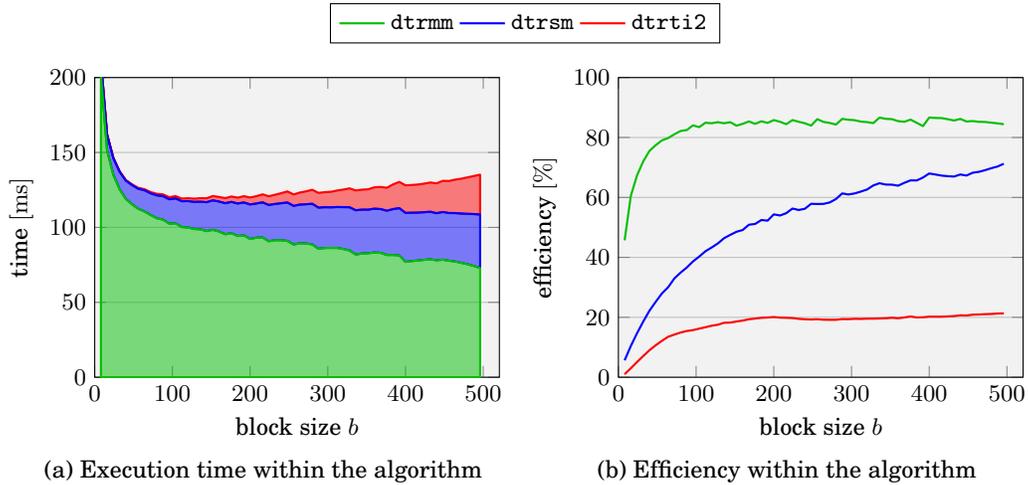

\parsum{analysis of effect of $b$ on performance}
The first two updates (\dtrmm\footnote{%
    \dtrmm: {\tt d}ouble precision {\tt tr}iangular {\tt m}atrix times {\tt
    m}atrix (\blasl3)
} and \dtrsm\footnote{%
    {\tt dtrsm}: {\tt d}ouble precision {\tt tr}iangular linear system {\tt
    s}olve with {\tt m}ultiple right-hand-sides (\blasl3)
}) of the algorithm in \autoref{fig:blocked_alg}, are calls to \blasl3 kernels,
while the last one (\dtrti2\footnote{%
    {\tt dtrti2}: {\tt d}ouble precision {\tt tr}iangular matrix {\tt i}nversion
    (unblocked \lapack)
}) invokes an unblocked algorithm based on \blasl1 and~2, which is equivalent to
$b = 1$.  For a matrix of size $n = 2000$, \autoref{fig:blocked_breakdown} gives
an idea \hyperref[fig:blocked_breakdown:time]{(a)} of how much these routines
contribute to the algorithm's total execution time and
\hyperref[fig:blocked_breakdown:eff]{(b)} what efficiency (with respect to the
processors theoretical compute bound) they operate at within the
algorithm.\footnote{%
    Executed on one core of an {\sc Intel Ivy~Bridge-EP E5-2680~v2} using {\sc
    OpenBLAS} for the \blas kernels \dtrmm and \dtrsm and reference \lapack for
    the unblocked \dtrti2.
}  For very small values of $b$, the compute intensity\footnote{%
    The ratio of floating point operations to memory operations.
} of \dtrmm and \dtrsm is so low that they are effectively memory bound, and
thus very inefficient.  As $b$ increases, the size of the three kernels grows
and so does their efficiency (see \autoref{fig:blocked_breakdown:eff}):
\dtrmm~(\ref*{plt:blocked_breakdown:dtrmm}) plateaus at \SI{85}{\percent} around
$b = 120$, \dtrsm's efficiency~(\ref*{plt:blocked_breakdown:dtrsm}) steadily
rises towards that of \dtrmm~(\ref*{plt:blocked_breakdown:dtrmm}), while
\dtrti2~(\ref*{plt:blocked_breakdown:dtrti2}) approaches its peak of only
\SI{20}{\percent} towards $b = 200$.  On the other side, when increasing $b$,
more and more computation is shifted from the \blasl3 routines to the low
performance \dtrti2~(\ref*{plt:blocked_breakdown:dtrti2}); beyond $b = 150$ this
low performance causes the overall runtime to increase.  This trade-off between
increasing \blasl3 performance and shifting the computation to a less efficient
unblocked kernel is a well known phenomenon inherent to all blocked algorithms.

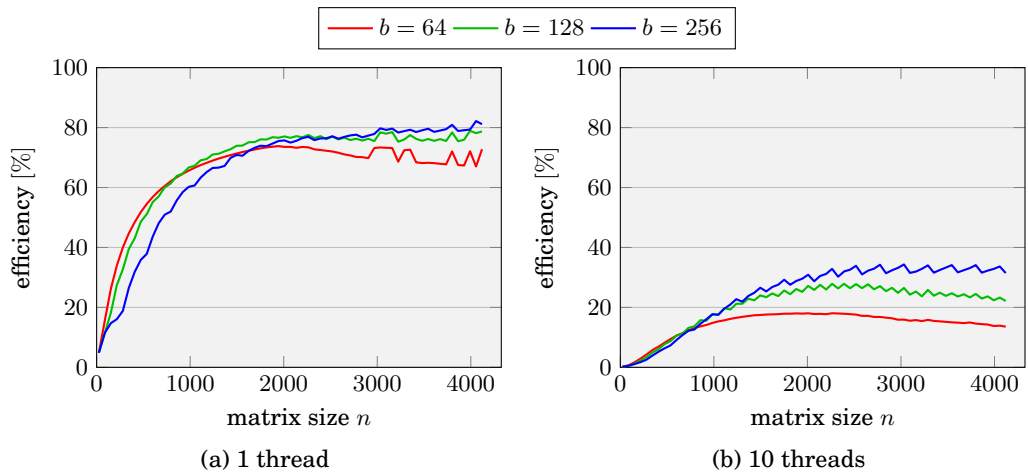
\begin{figure}[t]
    \centering\small

    \ref*{leg:blocked_bvalues}

    \begin{subfigure}{.5\textwidth}
        \tikzset{external/export=true}\tikzsetnextfilename{blocked_bvalues_1}
        \begin{tikzpicture}
            \begin{axis}[
                twocolplot,
                ymax=100,
                xlabel={matrix size $n$},
                ylabel={efficiency},
                y unit=\si\percent,
                legend to name=leg:blocked_bvalues,
                legend columns=-1,
            ]
                \addplot table[y=64] {figures/data/blocked_bvalues/eff1.dat};
                \addlegendentry{$b = 64$}
                \label{plt:blocked_bvalues:64}
                \addplot table[y=128] {figures/data/blocked_bvalues/eff1.dat};
                \addlegendentry{$b = 128$}
                \label{plt:blocked_bvalues:128}
                \addplot table[y=256] {figures/data/blocked_bvalues/eff1.dat};
                \addlegendentry{$b = 256$}
                \label{plt:blocked_bvalues:256}
            \end{axis}
        \end{tikzpicture}
        \caption{1~thread}
        \label{fig:blocked_bvalues:nt1}
    \end{subfigure}%
    \begin{subfigure}{.5\textwidth}
        \tikzset{external/export=true}\tikzsetnextfilename{blocked_bvalues_10}
        \begin{tikzpicture}
            \begin{axis}[
                twocolplot,
                ymax=100,
                xlabel={matrix size $n$},
                ylabel={efficiency},
                y unit=\si\percent,
            ]
                \addplot table[y=64] {figures/data/blocked_bvalues/eff10.dat};
                \addplot table[y=128] {figures/data/blocked_bvalues/eff10.dat};
                \addplot table[y=256] {figures/data/blocked_bvalues/eff10.dat};
            \end{axis}
        \end{tikzpicture}
        \caption{10~threads}
        \label{fig:blocked_bvalues:nt10}
    \end{subfigure}

    \caption{
        Efficiency of the blocked triangular inversion algorithm for different
        block sizes~$b$:  Different block sizes are optimal for different
        settings.
    }
    \label{fig:blocked_bvalues}
\end{figure}

\parsum{factors influencing the optimal $b$}
To further illustrate the importance of this trade-off,
\autoref{fig:blocked_bvalues} reports the performance of the presented algorithm
with 1 and 10 threads, for $b = 64, 128, 256$, and increasing matrix size $n$.
The performance results, which are representative for all blocked algorithms,
indicate that, both for different matrix sizes and different thread counts, the
ideal choice of $b$ varies.  In fact, the optimal choice among the three values
for single threaded \blas and matrix size $n = 500$ is $b =
64$~(\ref*{plt:blocked_bvalues:64}); however, at $n = 2000$, this choice would
be about \SI{10}{\percent} less efficient than $b =
256$~(\ref*{plt:blocked_bvalues:256}) on 1 core and only \SI{50}{\percent} as
fast on 10 cores.  Even though this trade-off is well known, it remains a
challenging and important optimization task when implementing and tuning any
blocked algorithm.

\subsection{Related Work}

\parsum{blocked algorithms}
\lapack \cite{lapack} is a well established library that provides blocked
algorithms for many higher-level operations, such as inversions, factorizations,
and reductions \cite{lapackblocked}.  These algorithm's block size $b$ is well
understood to be a crucial tuning factor
\cite{lapacktuning,hpltuning,factorizationtuning}. 

\parsum{algorithms by blocks}
An alternative to blocked algorithms worth mentioning are {\em
algorithms-by-blocks}, which are also known as {\em block algorithms} or {\em
tiled algorithms}.  These algorithms make use of shared memory systems in the
form of task-based parallelism.  To this end, most implementations not only
introduce a specialized storage scheme of matrices ``by block'', but also
propose specially tailored task scheduling algorithms.  Implementations of such
schedulers include {\sc QUARK} \cite{quark}, {\sc DAGuE} \cite{dague}, {\sc
SMPSs} \cite{smpssdla}, and {\sc SuperMatrix} \cite{supermatrix}.

}

%%% Local Variables:
%%% mode: latex
%%% TeX-master: "main"
%%% End:

    \section{Recursive Algorithms}
    \label{sec:recursive}
    {
\renewcommand\A{\drawmatrix[lower]A}
\newcommand\Atl{\drawmatrix[fill=plot2!40, lower]{A_{TL}}}
\newcommand\Abl{\drawmatrix[fill=plot3!40       ]{A_{BL}}}
\newcommand\Abr{\drawmatrix[fill=plot1!40, lower]{A_{BR}}}

Blocked algorithms can be translated into recursive algorithm by setting the
block size to $\frac n2$ and replacing the calls to the unblocked kernels with
recursive algorithm invocations.\footnote{%
    For a given operation, once the traversal direction is fixed, all blocked
    variants results in the same recursive algorithm.
}  Building on the same exemple operation used so far, this section details how
such recursive algorithms operate, and what factors influence their performance.

\begin{figure}[t]
    \centering\small

    \renewcommand\Atl{\drawmatrix[size=1.1, fill=plot2!40, lower]{A_{TL}}}
    \renewcommand\Abl{\drawmatrix[size=1.1, fill=plot3!40       ]{A_{BL}}}
    \renewcommand\Abr{\drawmatrix[size=1.1, fill=plot1!40, lower]{A_{BR}}}

    \subcaptionbox{%
        Recursive partitioning of $A$.
        \label{fig:recursive_alg:mat}
    }[.5\textwidth]{
        \centering
        \tikzset{external/export=true}\tikzsetnextfilename{recursive_alg_mat}
        \begin{tikzpicture}
            \filldraw[fill=plot2!40, draw=matrixborder!50!black] (0,         -0) --        ++(2.725, -2.725) coordinate[midway] (atl) -| (0,     -0);
            \filldraw[fill=plot3!40, draw=matrixborder!50!black] (0,     -2.775) rectangle ++(2.725, -2.725) node[black, midway] {$A_{BL}$};
            \filldraw[fill=plot1!40, draw=matrixborder!50!black] (2.775, -2.775) --        ++(2.725, -2.725) coordinate[midway] (abr) -| (2.775, -2.775);
            \foreach \shift in {0, 2.775}
                {\begin{scope}[shift={(\shift, -\shift)}, scale=.495454, opacity=.7]
                    \draw[matrixborder!50!black] (0, -2.75) -- ++(2.75, 0) -- ++(0, -2.75);
                    \foreach \shift in {0, 2.75}
                        {\begin{scope}[shift={(\shift, -\shift)}, scale=.5, opacity=.49]
                            \draw[matrixborder!50!black] (0, -2.75) -- ++(2.75, 0) -- ++(0, -2.75);
                            \foreach \shift in {0, 2.75}
                                {\begin{scope}[shift={(\shift, -\shift)}, scale=.5, opacity=.343]
                                    \draw[matrixborder!50!black] (0, -2.75) -- ++(2.75, 0) -- ++(0, -2.75);
                                    \foreach \shift in {0, 2.75}
                                        {\begin{scope}[shift={(\shift, -\shift)}, scale=.5, opacity=.2401]
                                            \draw[matrixborder!50!black] (0, -2.75) -- ++(2.75, 0) -- ++(0, -2.75);
                                        \end{scope}}
                                \end{scope}}
                        \end{scope}}
                \end{scope}}

            \draw[decorate, decoration=brace] (-.05,   -5.5) -- ++(0,        5.5) node[midway, anchor=east]  {$n$};
            \draw[decorate, decoration=brace] (5.5,   -5.55) -- ++(-5.5,       0) node[midway, anchor=north] {$n$};
            \draw[decorate, decoration=brace] (2.775,     0) -- ++(0,     -2.725) node[midway, anchor=west]  {$\frac{n}2$};
            \draw[decorate, decoration=brace] (0,       .05) -- ++(2.725,      0) node[midway, anchor=south] {$\frac{n}2$};

            \node at (atl) {$A_{TL}$};
            \node at (abr) {$A_{BR}$};
        \end{tikzpicture}
        \tikzset{external/export=false}
    }%
    \subcaptionbox{%
        Updates performed at each recursion level. \dtrtrir: recursion if
        $\frac n2 > c$, \dtrti2 otherwise.
        \label{fig:recursive_alg:ops}
    }[.5\textwidth]{
        \setlength\tabcolsep{0pt}
        \begin{tabular}{ll}
            (1) \dtrtrir:   &\ $\Atl \coloneqq \Atl^{-1}        \vphantom{\drawmatrix[height=1.2]X}$\\
            (2) \dtrmm:     &\ $\Abl \coloneqq \Abl\;\Atl       \vphantom{\drawmatrix[height=1.2]X}$\\
            (3) \dtrsm:     &\ $\Abl \coloneqq -\Abr^{-1} \Abl  \vphantom{\drawmatrix[height=1.2]X}$\\
            (4) \dtrtrir:   &\ $\Abr \coloneqq \Abr^{-1}        \vphantom{\drawmatrix[height=1.2]X}$\\
        \end{tabular}

        \vspace{.5cm}
    }

    \vspace{-\medskipamount}

    \caption{%
        Recursive algorithm for the inversion of a lower triangular matrix.
    }
    \label{fig:recursive_alg}
\end{figure}
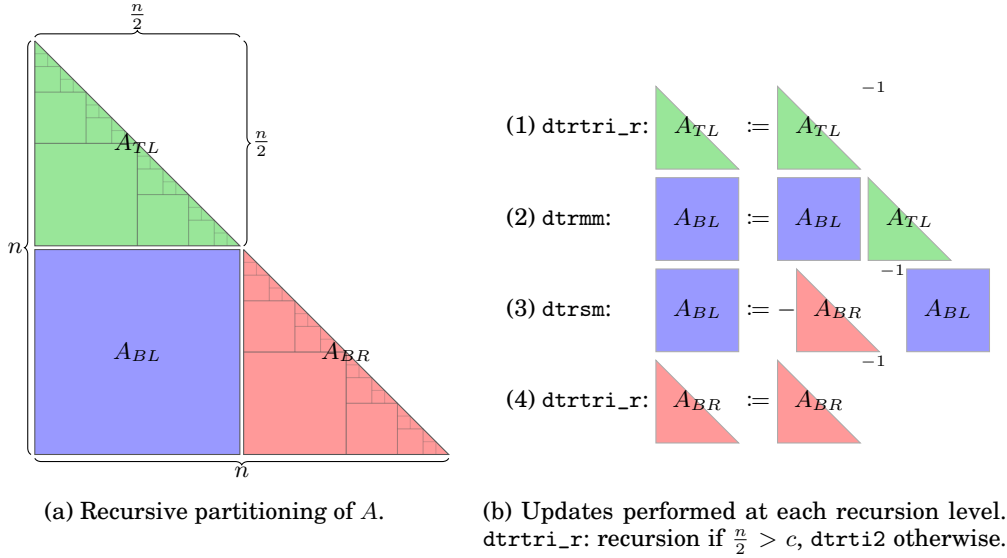

\parsum{Recursive algorithm in detail}
As shown in \autoref{fig:recursive_alg:mat}, the recursive algorithm for this
operation starts off by splitting both dimensions of the lower triangular $n
\times n$ input matrix $\A$ in half,\footnote{%
    For alignment, we ensure that, where possible,  the sizes of the
    sub-matrices are multiples of \num8, allowing the two ``halves'' to be of
    slightly different sizes.
} exposing the quadrants $\Atl$, $\Abl$, and $\Abr$; the updates in
\autoref{fig:recursive_alg:ops} are then applies to the quadrants:  Two \blasl3
invocations surrounded by two recursive applications of the inversion algorithm
to $\Atl$ and $\Abr$.\footnote{%
    To increase performance by \SI5{\percent} to \SI{15}{\percent} (depending on
    the matrix size), one could first invert $A_{BR}$ and replace the linear
    system solve (\dtrsm) by a more efficient matrix-matrix multiplication
    (\dtrmm). Since this change makes the algorithm numerically unstable
    \cite{trtristability}, we do not include results.
}  In principle, one could continue the recursion down to
the scalar matrix entries, where the operation turns into a trivial $a \coloneqq
1/a$; however, in order to minimize the number of tiny \blasl3 invocations, we
introduce the {\em crossover size} $c$:\footnote{%
    In \cite{dlarec} it is referred to as a ``blocking factor''.
} whenever $n < c$ the
algorithm switches to \lapack's unblocked \dtrti2.

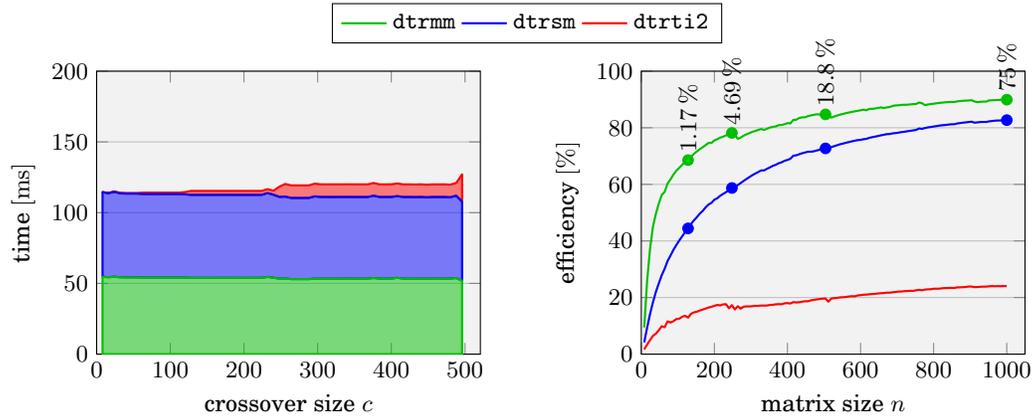
\begin{figure}[t]
    \centering\small

    \ref*{leg:recursive_breakdown}

    \vspace{-\medskipamount}

    \tikzset{external/export=true}
    \begin{subfigure}[t]{.48\textwidth}
        \tikzsetnextfilename{recursive_breakdown_time}
        \begin{tikzpicture}
            \begin{axis}[
                twocolplot,
                ymax=200,
                xlabel={crossover size $c$},
                ylabel={time},
                y unit=\si\ms,
                stack plots=y,
            ]
                \addplot[plot2, fill, fill opacity=.5] table[y=dtrmm]  {figures/data/recursive_breakdown/time.dat} \closedcycle;
                \addplot[plot3, fill, fill opacity=.5] table[y=dtrsm]  {figures/data/recursive_breakdown/time.dat} \closedcycle;
                \addplot[plot1, fill, fill opacity=.5] table[y=dtrti2] {figures/data/recursive_breakdown/time.dat} \closedcycle;
            \end{axis}
        \end{tikzpicture}
        \caption{%
            Execution time within the recursive inversion algorithm with
            increasing crossover size~$c$.
        }
        \label{fig:recursive_breakdown:time}
    \end{subfigure}
    \hfill
    \begin{subfigure}[t]{.48\textwidth}
        \tikzsetnextfilename{recursive_breakdown_eff}
        \begin{tikzpicture}
            \begin{axis}[
                twocolplot,
                ymax=100,
                xlabel={matrix size $n$},
                ylabel={efficiency},
                y unit=\si\percent,
                legend to name=leg:recursive_breakdown,
                legend columns=-1,
            ]
                \addplot[name path=dtrmm, plot2] table[x=n, y=dtrmm]
                {figures/data/recursive_breakdown_parts/eff.dat};
                \addlegendentry{\dtrmm}
                \addplot[name path=dtrsm, plot3] table[x=n, y=dtrsm]
                {figures/data/recursive_breakdown_parts/eff.dat};
                \addlegendentry{\dtrsm}
                \addplot[name path=dtrti2, plot1] table[x=n, y=dtrti2]
                {figures/data/recursive_breakdown_parts/eff.dat};
                \addlegendentry{\dtrti2}

                \path[name path=1000] (axis cs: 1000,0) -- (axis cs:1000,100);
                \filldraw[name intersections={of=dtrmm and 1000}, plot2]
                (intersection-1) circle (2pt) coordinate (dtrmm 1000);
                \filldraw[name intersections={of=dtrsm and 1000}, plot3]
                (intersection-1) circle (2pt) coordinate (dtrsm 1000);

                \path[name path=500] (axis cs: 504,0) -- (axis cs:504,100);
                \filldraw[name intersections={of=dtrmm and 500}, plot2]
                (intersection-1) circle (2pt) coordinate (dtrmm 500);
                \filldraw[name intersections={of=dtrsm and 500}, plot3]
                (intersection-1) circle (2pt) coordinate (dtrsm 500);

                \path[name path=250] (axis cs: 248,0) -- (axis cs:248,100);
                \filldraw[name intersections={of=dtrmm and 250}, plot2]
                (intersection-1) circle (2pt) coordinate (dtrmm 250);
                \filldraw[name intersections={of=dtrsm and 250}, plot3]
                (intersection-1) circle (2pt) coordinate (dtrsm 250);

                \path[name path=125] (axis cs: 128,0) -- (axis cs:128,100);
                \filldraw[name intersections={of=dtrmm and 125}, plot2]
                (intersection-1) circle (2pt) coordinate (dtrmm 125);
                \filldraw[name intersections={of=dtrsm and 125}, plot3]
                (intersection-1) circle (2pt) coordinate (dtrsm 125);
            \end{axis}
            \foreach \n/\p in {1000/75, 500/18.8, 250/4.69, 125/1.17}
                \path (dtrmm \n) node[anchor=west, rotate=90]
                    {\SI{\p}{\percent}};
        \end{tikzpicture}
        \caption{%
            Efficiency for square matrices.  Annotations: \si{\percent} of FLOPs
            in the inversion covered by the \dtrsm and \dtrmm (assuming $c <
            100$).
        }
        \label{fig:recursive_breakdown:eff}
    \end{subfigure}
    \caption{%
        Breakdown of the recursive inversion of a matrix of size $n = 2000$.
    }
    \label{fig:recursive_breakdown}
\end{figure}

\parsum{analysis of effect of $c$ on performance}
In contrast to the block size $b$ of blocked algorithms, which requires careful
tuning for high performance reasons, the choice of the crossover size $c$ is
entirely straightforward.  Indeed, as \autoref{fig:recursive_breakdown:time}
suggests, as long as $c$ is kept small, the overall performance is only
moderately affected; the reason becomes clear by inspecting the recursive
algorithm: Changes in $c$ have no effect whatsoever on the large \blas calls,
whose size is solely determined by $n$.  In fact, considering that the inversion
of a triangular $n \times n$ matrix takes $\frac{n^3}3$ floating point
operations (FLOPs), the two \blasl3 calls in the top recursion level (\dtrmm and
\dtrsm), which together cover $2 \bigl(\frac n2 \bigr)^3$ FLOPs, account for
$$
    \frac{2 \bigl(\frac n2\bigr)^3}{\frac{n^3}3} =
    \frac{\frac{n^3}4}{\frac{n^3}3} =
    \frac34 =
    \SI{75}{\percent}
$$
of the algorithm's entire FLOPs.  This is visualized in
\autoref{fig:recursive_breakdown:eff};  assuming $c < 100$, out of the total
FLOPs for the inversion, the \dtrmm and \dtrsm account for over
\SI{99.6}{\percent}:  \SI{75}{\percent} of the FLOPs are covered by the two
calls on the first recursion level performing at about \SI{86}{\percent}
efficiency; on the second level, \SI{18.8}{\percent} perform at
\SI{79}{\percent}, on the third, \SI{4.69}{\percent} at \SI{68}{\percent}, and
on the fourth, \SI{1.17}{\percent} at \SI{57}{\percent}.  As a result, only
$< \SI{.4}{\percent}$ of the FLOPs attain an efficiency below \SI{57}{\percent}.

\parsum{choosing $c$}
This analysis confirms that a small crossover size does not harm the performance
of the \blasl3 kernels.  Moreover, \autoref{fig:recursive_breakdown:time}
provides evidence that choosing $c$ as small as \num8 does not cause any
performance penalty.  A comparison of recursive algorithms with \lapack's
unblocked kernels has shown that, the unblocked kernel is slightly faster than
the recursive algorithm for small matrices within the processor's L1~cache.
Hence, for the remainder of this paper, we use $c = 24$.

\subsection{Related Work}
In a series of works \cite{cacheoblivious1,cacheoblivious2}, recursive
algorithms---coined as ``cache-oblivious algorithms''---were proven to be
optimal in the sense that they minimize data movement independently of cache
sizes.  Recursion as an alternative to \lapack's blocked algorithms has been
proposed in several publications: Starting from the description of recursive
versions of the Cholesky and LU decompositions in \cite{dlarec}, FORTRAN~90
implementations for these operations are developed in \cite{cholrec} and
\cite{lurec}, specialized recursive storage schemes are proposed by
\cite{recstorage} and are applied to matrix decompositions in
\cite{lawra,cholrecstorage}.  Recursion is applied to the QR decomposition in
\cite{qrrec} and triangular matrix inversion in \cite{trinvrec}.  Finally,
several Sylvester-type equation solvers are implemented with recursive
algorithms in the RECSY library \cite{recsy}.  While many of these works
describe their techniques in much detail, they each only consider one or a very
limited set of operations---until now, no comprehensive implementation of
recursive algorithms comparable to \lapack's range of blocked routines is
available.

\parsum{recursive Cholesky in \lapack}
\lapack's most recent release (version~3.6.0) introduces a recursive version of
the Cholesky factorization, named {\tt dpotrf2}; this routine, which is fully
recursive down to the scalar level is used in \lapack's blocked \dpotrf as a
replacement for the unblocked \dpotf2.

}

%%% Local Variables:
%%% mode: latex
%%% TeX-master: "main"
%%% End:

    \section{Perfectly Tuned Blocked vs. Recursive Algorithms}
    \label{sec:bvsr}
    {
\renewcommand\A{\drawmatrix A}
\renewcommand\L{\drawmatrix[lower]L\!}
\newcommand\LT{\!\drawmatrix[upper]{\hphantom{^T}L^T}}
\newcommand\U{\!\drawmatrix[upper]U}
\renewcommand\P{\drawmatrix[dashed]P}

\parsum{general setup}
In this section, we compare recursive and blocked algorithms in terms of
performance.  For the former, we chose a reasonable crossover size at $c = 24$,
and kept it constant; for the latter, we undertook a extensive tuning process,
for each problem size, timing all the algorithmic variants for all reasonable
block sizes .  While in practice such a process is clearly infeasible, we
carried it out to establish a strict upper bound to the best possible
performance achievable by blocked algorithms.  The comparison was performed on a
10-core {\sc Intel IvyBridge E5-2680~v2} processor running at \SI{2.8}{\giga\Hz}
(Turbo Boost: \SI{3.6}{\giga\Hz}), using {\sc OpenBLAS} (version~0.2.15)
\cite{openblas} for the \blas kernels, and reference \lapack (version~3.5) for
the unblocked calls.  A comparison against tuned \lapack implementations is
presented in \autoref{sec:vslibs}.

\begin{figure}[t]
    \centering\small

    \ref*{leg:bvsr_trinv}

    \vspace\medskipamount

    \begin{subfigure}{.5\textwidth}
        \tikzset{external/export=true}\tikzsetnextfilename{bvsr_trinv_1}
        \begin{tikzpicture}
            \begin{axis}[
                twocolplot,
                ymax=100,
                xlabel={matrix size $n$},
                ylabel={efficiency},
                y unit=\si\percent,
                legend to name=leg:bvsr_trinv,
                legend columns=5,
            ]
                \addlegendimage{empty legend} \addlegendentry{recursive algorithm:}
                \addlegendimage{plot1}
                \addlegendentry{}
                \label{plt:bvsr_trinv:r}
                \addlegendimage{empty legend} \addlegendentry{}
                \addlegendimage{empty legend} \addlegendentry{}
                \addlegendimage{empty legend} \addlegendentry{}
                \addlegendimage{empty legend} \addlegendentry{blocked algorithms:}
                \addplot[plot2] table[y=dtrtri_b1] {figures/data/bvsr/dtrtri1.dat};
                \addlegendentry{1}
                \label{plt:bvsr_trinv:b1}
                \addplot[plot3] table[y=dtrtri_b2] {figures/data/bvsr/dtrtri1.dat};
                \addlegendentry{2}
                \label{plt:bvsr_trinv:b2}
                \addplot[plot4] table[y=dtrtri_b3] {figures/data/bvsr/dtrtri1.dat};
                \addlegendentry{3}
                \label{plt:bvsr_trinv:b3}
                \addplot[plot5] table[y=dtrtri_b4] {figures/data/bvsr/dtrtri1.dat};
                \addlegendentry{4}
                \label{plt:bvsr_trinv:b4}
                \addlegendimage{empty legend} \addlegendentry{reference \lapack:}
                \addplot[gray] table[y=lapack] {figures/data/bvsr/dtrtri1.dat};
                \addlegendentry{}
                \label{plt:bvsr_trinv:lapack}
                \addplot[white, double=plot1] table[y=dtrtri_r] {figures/data/bvsr/dtrtri1.dat};
            \end{axis}
        \end{tikzpicture}
        \caption{1~thread}
        \label{fig:bvsr_trinv:1}
    \end{subfigure}%
    \begin{subfigure}{.5\textwidth}
        \tikzset{external/export=true}\tikzsetnextfilename{bvsr_trinv_10}
        \begin{tikzpicture}
            \begin{axis}[
                twocolplot,
                ymax=100,
                xlabel={matrix size $n$},
                ylabel={efficiency},
                y unit=\si\percent,
            ]
                \addplot[plot2] table[y=dtrtri_b1] {figures/data/bvsr/dtrtri10.dat};
                \addplot[plot3] table[y=dtrtri_b2] {figures/data/bvsr/dtrtri10.dat};
                \addplot[plot4] table[y=dtrtri_b3] {figures/data/bvsr/dtrtri10.dat};
                \addplot[plot5] table[y=dtrtri_b4] {figures/data/bvsr/dtrtri10.dat};
                \addplot[gray] table[y=lapack] {figures/data/bvsr/dtrtri10.dat};
                \addplot[plot1] table[y=dtrtri_r] {figures/data/bvsr/dtrtri10.dat};
            \end{axis}
        \end{tikzpicture}
        \caption{10~threads}
        \label{fig:bvsr_trinv:10}
    \end{subfigure}

    \caption{
        Blocked algorithms (optimal $b$) vs. recursive algorithm ($c = 24$) for
        the inversion of a lower triangular matrix (\lapack: \dtrtri).
    }
    \label{fig:bvsr_trinv}
\end{figure}
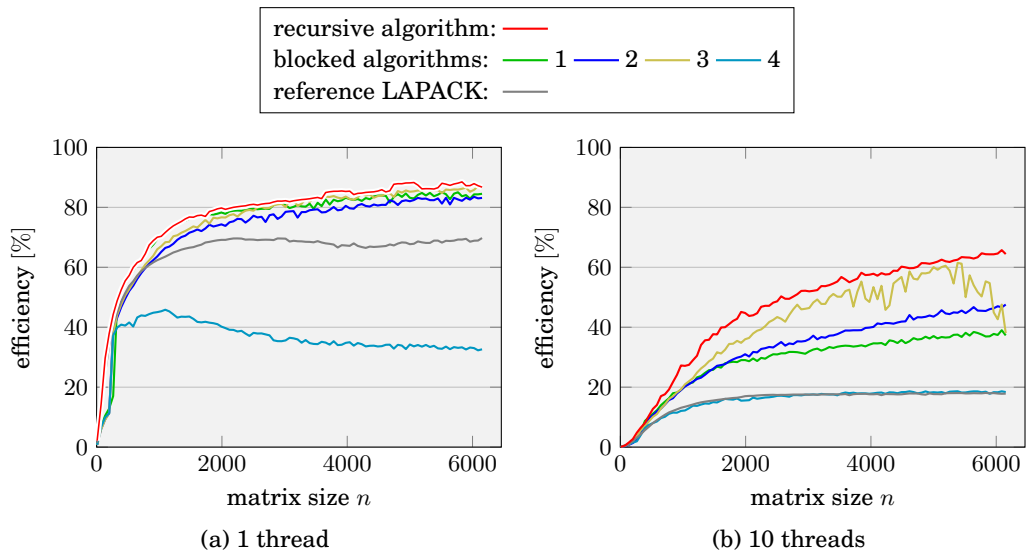

\parsum{analysis: triangular inversion}
\hyperref[fig:bvsr_trinv]{Figure~\ref*{fig:bvsr_trinv}} reports the performance
of the four unblocked algorithms and the recursive algorithm for the inversion
of a triangular matrix.  For single-threaded {\sc OpenBLAS}, the recursive
algorithm~(\ref*{plt:bvsr_trinv:r}) is on par with or slightly more efficient
than the best blocked algorithm (which changes from
variant~1~(\ref*{plt:bvsr_trinv:b1}) to variant~3~(\ref*{plt:bvsr_trinv:b3})
around $n = 3000$); when using all 10~cores of the CPU, the recursive algorithm
consistently outperforms all blocked algorithms.  We stress that these results
come at the cost of expensive tuning for the blocked algorithms, and with no
optimization at all for the recursive ones.  In comparison, reference
\lapack~(\ref*{plt:bvsr_trinv:lapack}) with default block size $b = 64$ is with
1 and 10~cores, respectively, about \SI{15}{\percent} and \SI{45}{\percent}
slower than the recursive algorithm.

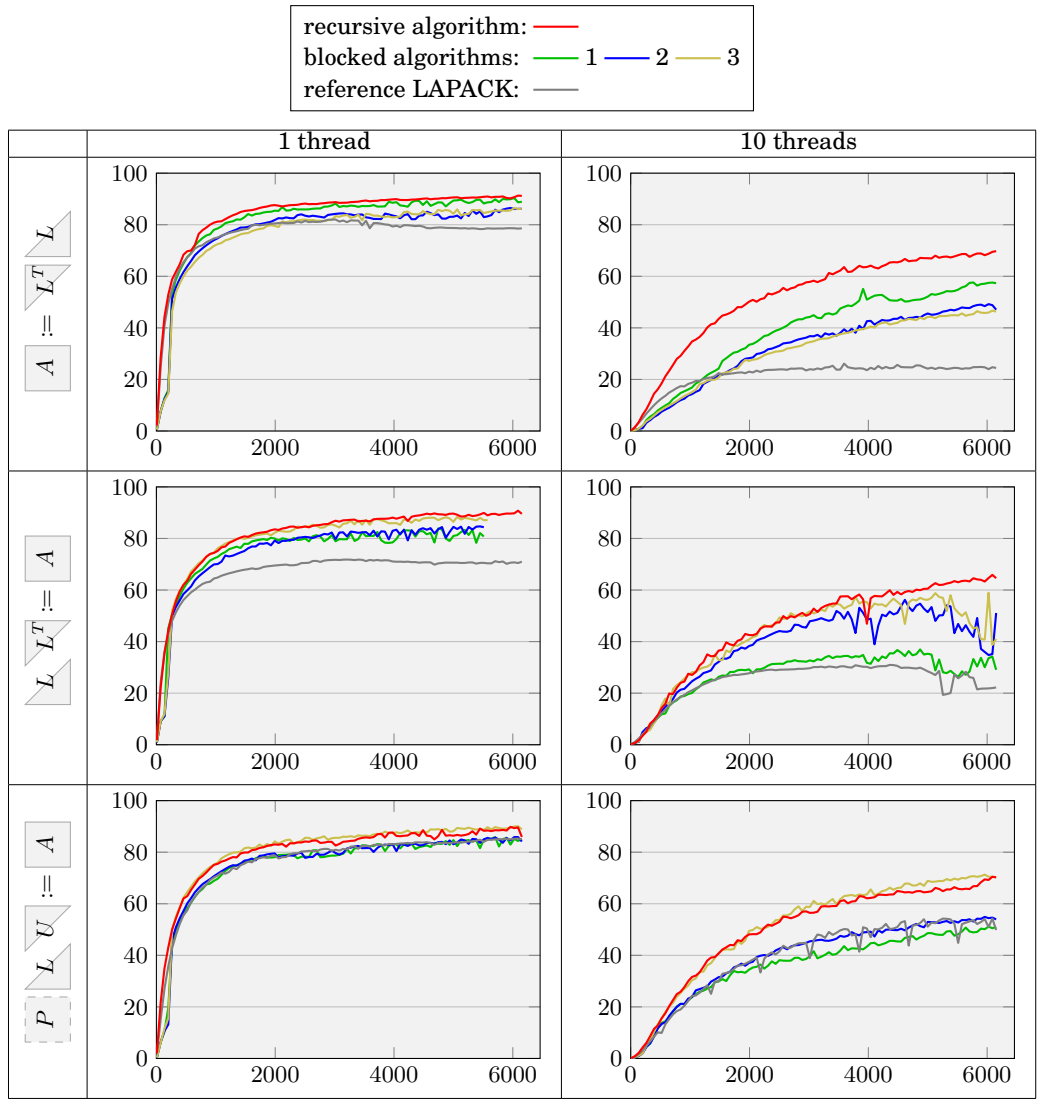
\begin{figure}[t]
    \centering\small

    \drawmatrixset{size=.6}

    \pgfplotsset{
        every axis/.append style={
            height=.36\textwidth,
            width=.48\textwidth,
            ymax=100,
            label shift=-1cm,
        },
        every axis plot/.append style={unbounded coords=jump},
        recplot/.style={white, double=plot1}
    }

    \ref*{leg:bvsr_others}

    \vspace\medskipamount
    \begin{tabular}{|c|c|c|}
        \hline
        &1~thread &10~threads \\
        \hline
        \rotatebox{90}{\qquad \quad $\A \coloneqq \LT \; \L$}
        &%
        \tikzset{external/export=true}\tikzsetnextfilename{bvsr_others_dlauum_1}%
        \begin{tikzpicture}
            \begin{axis}[
                legend to name=leg:bvsr_others,
                legend columns=4,
            ]
                \addlegendimage{empty legend} \addlegendentry{recursive algorithm:}
                \addlegendimage{plot1}
                \addlegendentry{}
                \addlegendimage{empty legend} \addlegendentry{}
                \addlegendimage{empty legend} \addlegendentry{}
                \addlegendimage{empty legend} \addlegendentry{blocked algorithms:}
                \addplot[plot2] table[y=dlauum_b1] {figures/data/bvsr/dlauum1.dat};
                \addlegendentry{1}
                \addplot[plot3] table[y=dlauum_b2] {figures/data/bvsr/dlauum1.dat};
                \addlegendentry{2}
                \addplot[plot4] table[y=dlauum_b3] {figures/data/bvsr/dlauum1.dat};
                \addlegendentry{3}
                \addlegendimage{empty legend} \addlegendentry{reference \lapack:}
                \addplot[gray]  table[y=lapack]    {figures/data/bvsr/dlauum1.dat};
                \addlegendentry{}
                \addplot[plot1] table[y=dlauum_r]  {figures/data/bvsr/dlauum1.dat};
            \end{axis}
        \end{tikzpicture}
        &%
        \tikzset{external/export=true}\tikzsetnextfilename{bvsr_others_dlauum_10}%
        \begin{tikzpicture}
            \begin{axis}
                \addplot[plot2] table[y=dlauum_b1] {figures/data/bvsr/dlauum10.dat};
                \addplot[plot3] table[y=dlauum_b2] {figures/data/bvsr/dlauum10.dat};
                \addplot[plot4] table[y=dlauum_b3] {figures/data/bvsr/dlauum10.dat};
                \addplot[gray]  table[y=lapack]    {figures/data/bvsr/dlauum10.dat};
                \addplot[plot1] table[y=dlauum_r]  {figures/data/bvsr/dlauum10.dat};
            \end{axis}
        \end{tikzpicture} \\
        \hline
        \rotatebox{90}{\qquad \quad $\L \LT \coloneqq \A$}
        &%
        \tikzset{external/export=true}\tikzsetnextfilename{bvsr_others_dpotrf_1}%
        \begin{tikzpicture}
            \begin{axis}
                \addplot[plot2] table[y=dpotrf_b1] {figures/data/bvsr/dpotrf1.dat};
                \addplot[plot3] table[y=dpotrf_b2] {figures/data/bvsr/dpotrf1.dat};
                \addplot[plot4] table[y=dpotrf_b3] {figures/data/bvsr/dpotrf1.dat};
                \addplot[gray]  table[y=lapack]    {figures/data/bvsr/dpotrf1.dat};
                \addplot[plot1] table[y=dpotrf_r]  {figures/data/bvsr/dpotrf1.dat};
            \end{axis}
        \end{tikzpicture}
        &%
        \tikzset{external/export=true}\tikzsetnextfilename{bvsr_others_dpotrf_10}%
        \begin{tikzpicture}
            \begin{axis}
                \addplot[plot2] table[y=dpotrf_b1] {figures/data/bvsr/dpotrf10.dat};
                \addplot[plot3] table[y=dpotrf_b2] {figures/data/bvsr/dpotrf10.dat};
                \addplot[plot4] table[y=dpotrf_b3] {figures/data/bvsr/dpotrf10.dat};
                \addplot[gray]  table[y=lapack]    {figures/data/bvsr/dpotrf10.dat};
                \addplot[plot1] table[y=dpotrf_r]  {figures/data/bvsr/dpotrf10.dat};
            \end{axis}
        \end{tikzpicture} \\
        \hline
        \rotatebox{90}{\qquad $\P \; \L \U \coloneqq \A$}
        &%
        \tikzset{external/export=true}\tikzsetnextfilename{bvsr_others_dgetrf_1}%
        \begin{tikzpicture}
            \begin{axis}
                \addplot[plot2] table[y=dgetrf_b1] {figures/data/bvsr/dgetrf1.dat};
                \addplot[plot3] table[y=dgetrf_b2] {figures/data/bvsr/dgetrf1.dat};
                \addplot[plot4] table[y=dgetrf_b3] {figures/data/bvsr/dgetrf1.dat};
                \addplot[gray]  table[y=lapack]    {figures/data/bvsr/dgetrf1.dat};
                \addplot[plot1] table[y=dgetrf_r]  {figures/data/bvsr/dgetrf1.dat};
            \end{axis}
        \end{tikzpicture}
        &%
        \tikzset{external/export=true}\tikzsetnextfilename{bvsr_others_dgetrf_10}%
        \begin{tikzpicture}
            \begin{axis}
                \addplot[plot2] table[y=dgetrf_b1] {figures/data/bvsr/dgetrf10.dat};
                \addplot[plot3] table[y=dgetrf_b2] {figures/data/bvsr/dgetrf10.dat};
                \addplot[plot4] table[y=dgetrf_b3] {figures/data/bvsr/dgetrf10.dat};
                \addplot[gray] table[y=lapack] {figures/data/bvsr/dgetrf10.dat};
                \addplot[plot1] table[y=dgetrf_r] {figures/data/bvsr/dgetrf10.dat};
            \end{axis}
        \end{tikzpicture} \\
        \hline
    \end{tabular}
    \caption{%
        Blocked algorithms (optimal $b$) vs. recursive algorithm ($c = 24$).\\
        $x$-axes: matrix size $n$; $y$-axes: efficiency~[\si\percent].
    }
    \label{fig:bvsr_others}
\end{figure}

%%% Local Variables:
%%% mode: latex
%%% TeX-master: "../main"
%%% End:

\parsum{other matrix operations}
\hyperref[fig:bvsr_others]{Figure~\ref*{fig:bvsr_others}} presents results for
the following operations:
\begin{itemize}
    \item Multiplication of a triangular matrix with its transpose from the left
        $\A \coloneqq \LT \; \L$ \\
        (\lapack: \dlauum),
    \item Cholesky decomposition of a symmetric positive definite matrix $\L \LT
        \coloneqq \A$ \\
        (\lapack: \dpotrf), and
    \item LU decomposition of a square matrix with full pivoting $\P \; \L \U
        \coloneqq \A$ \\
        (\lapack: \dgetrf)
\end{itemize}
The results are consistent across the board: the unoptimized recursive algorithm
is always comparable with (or faster than) the fastest blocked algorithm with a
heavily optimized block size.  We performed similar comparisons on {\sc Intel
Sandy Bridge} and {\sc Haswell} processors, linking to both MKL and {\sc
OpenBLAS}; in all cases, the results are in line with those reported here.

}

%%% Local Variables:
%%% mode: latex
%%% TeX-master: "main"
%%% End:

    \section{ReLAPACK}
    \label{sec:relapack}
    {
% some commands to simplify equations
\renewcommand\A{\drawmatrix A}
\newcommand\B{\drawmatrix B}
\newcommand\C{\drawmatrix C}
\renewcommand\L{\drawmatrix[lower]L\!}
\newcommand\LT{\!\drawmatrix[upper]{\hphantom{^H}L^H}}
\newcommand\Linv{\!\!\!\!\drawmatrix[lower]{\hphantom{^{-1}}L^{-1}}\!}
\newcommand\x{\drawmatrix[width=0]x}
\newcommand\D{\!\drawmatrix[diag]D\!}
% LU stuff
\newcommand\lP{\drawmatrix[dashed]P}
\newcommand\lPT{\drawmatrix[dashed]{\hphantom{^H}P^H}}
\newcommand\lL{\drawmatrix[width=.4, lower]L}
\newcommand\lU{\drawmatrix[size=.4, bbox height=.6, upper]U}
\newcommand\lA{\drawmatrix[width=.4]A}
% LDL stuff
\newcommand\lB{\drawmatrix[height=.4, bbox height=.6]B}
\newcommand\lAT{\drawmatrix[height=.4, bbox height=.6]{\hphantom{^H}A^H}}
\newcommand\Dsmall{\drawmatrix[diag, size=.4, bbox height=.6]D}
% Sylvester stuff
\newcommand\sA{\!\drawmatrix[upper]A}
\newcommand\sB{\drawmatrix[size=.4, bbox height=.6, upper]B}
\newcommand\sC{\drawmatrix[width=.4]C}
\newcommand\sD{\!\drawmatrix[upper]D}
\newcommand\sE{\drawmatrix[size=.4, bbox height=.6, upper]E}
\newcommand\sF{\drawmatrix[width=.4]F}
\newcommand\sX{\drawmatrix[width=.4]X}
\newcommand\sR{\drawmatrix[width=.4]R}
\newcommand\sL{\drawmatrix[width=.4]L}
% QR stuff
\newcommand\qQ{\drawmatrix[width=.4]Q}
\newcommand\qR{\drawmatrix[upper, size=.4, bbox height=.6]R}
\newcommand\qA{\drawmatrix[width=.4]A}
% reduction stuff
\newcommand\rQ{\drawmatrix Q}
\newcommand\rQT{\drawmatrix{\hphantom{^H}Q^H}}
\newcommand\rPT{\drawmatrix{\hphantom{^H}P^H}}
\newcommand\rT{\drawmatrix[banded, bandwidth=.05]T}
\newcommand\rB{\drawmatrix[banded, bandwidth=.05, lower]B}
\newcommand\rH{\drawmatrix[lower banded, lower bandwidth=.05]H}
% banded stuff
\newcommand\bL{\drawmatrix[banded, lower]L\!}
\newcommand\bLT{\!\drawmatrix[banded, upper]{\hphantom{^H}L^H}}
\newcommand\bU{\!\drawmatrix[banded, bandwidth=.2, upper]U}
\newcommand\bA{\drawmatrix[banded]A}
\newcommand\bAdiff{\drawmatrix[banded, upper bandwidth=.2]A}

\parsum{\relapack introduction}
We have established that the tedious and time-consuming tuning process, which is
indispensable to attain close-to-optimal performance with blocked algorithms,
can be avoided for recursive algorithms without sacrificing performance.  In
this section, we present {\sc Recursive \lapack Collection} (\relapack), an
open-source library of \lapack operations implemented in a purely recursive
fashion.\footnote{%
    \relapack is available on GitHub: \url{http://github.com/HPAC/ReLAPACK}.
}  All operations are available in the four standard data types ($\texttt{\it x}
\in \{\texttt s, \texttt d, \texttt c, \texttt z\}$) and, since  \lapack's
interface is preserved, \relapack can be employed effortlessly in existing
codes.  Below, we list the operations currently included in \relapack:

\begin{itemize}
    \item \phantomsection\xroutine{lauum}:\label{routine:xlauum}
        Multiplication of a triangular matrix with its (complex conjugate)
        transpose, resulting in a symmetric (Hermitian) matrix; example:
        $\A \coloneqq \LT \; \L$. This operation arises as part of the inversion
        of symmetric positive definite matrices (\xroutine{potrf}).

    \item \phantomsection\xroutine[sy]{gst}:\label{routine:xsygst}%
        \footnote{%
            $\texttt{\it xsy} \in \{\texttt{ssy}, \texttt{dsy}, \texttt{che},
            \texttt{zhe}\}$, i.e., the complex variants are called {\tt chegst}
            and {\tt zhegst}.
        } Simultaneous two-sided multiplication of a symmetric (Hermitian)
        matrix with a triangular matrix and its (complex conjugate) transpose
        example: $\A \coloneqq \LT \; \A \; \L$.  This routines, which also
        cover the case where instead of $\L$ its inverse $\Linv$ is applied, is
        used to reduce symmetric (Hermitian) positive definite generalized
        eigenvalue problems (e.g., $\A \, \x = \lambda \, \B \, \x$) to the
        standard form ($\A \, \x = \lambda \x$), for instance in
        \xroutine[sy]{gv}.

        {\em User Note}:
        \relapack's \xroutine{sygst} performs about \SI{30}{\percent} fewer
        FLOPs than \lapack's blocked algorithm by internally using a temporary
        buffer of size $\frac n2 \times \frac n2$.  This algorithmic improvement
        over LAPACK, which is inspired by the blocked algorithms for this
        operation in of the {\sc libFLAME} library \cite{flame}, can optionally
        be disabled to avoid the memory overhead.

    \item \phantomsection\xroutine{trtri}:\label{routine:xtrtri}
        Inversion of a triangular matrix; example: $\L \coloneqq \Linv$.  This
        routine serves as a building block in the inversions of general matrices
        (\xroutine{getri}), and symmetric positive definite matrices
        (\xroutine{potri}).

    \item \phantomsection\xroutine{potrf}:\label{routine:xpotrf}
        Cholesky decomposition of a symmetric (Hermitian) positive definite
        matrix: $\L \LT \coloneqq \A$.  This routine is the central
        building block for many operations on such matrices, for instance:
        inversion (\xroutine{potri}), solution of linear systems
        (\xroutine{posv}), and reduction of generalized eigenvalue problems to
        standard form (\xroutine[sy]{gv}).

    \item \phantomsection\xroutine[sy]{trf}:\label{routine:xsytrf}%
        \footnote{%
            In complex arithmetic, there are routines for both symmetric ({\tt
            csytrf} and {\tt zsytrf}) and Hermitian matrices ({\tt chetrf} and
            {\tt zhetrf}).
        }
        LDL decomposition of a symmetric (or Hermitian) matrix; example: $\L \D
        \LT \coloneqq \A$.  This routine is used with symmetric indefinite
        matrices for inversions (\xroutine[sy]{tri}), and solutions of linear
        systems (\xroutine[sy]{sv}).

        {\em User Note}:
        In contrast to \lapack's \xroutine[sy]{trf}, which requires a temporary
        buffer of size $n \times b$, \relapack requires a buffer of size $T = n
        \times \frac n2$.  Conforming to the signature of \lapack, the
        works-space querying mechanism (via {\tt lwork = -1}) reports the
        required size $T$ and the buffer is expected as the {\tt work} argument.
        However, to avoid conflicts when plugging \relapack into existing codes,
        should the passed auxiliary buffer be to small, \xroutine[sy]{trf}
        allocates (and frees) such a buffer on its own.

        {\em Implementation Note}:
        Since \lapack's interface for \xroutine[sy]{trf}'s unblocked building
        block \xroutine{la{\it sy}f} is not directly suitable for recursion,
        \relapack's \xroutine[sy]{trf} comes with two auxiliary routines:  An
        unblocked kernel, that is a slight modification of \lapack's
        \xroutine{la{\it sy}f} and, since \blas does not support symmetric
        operations of the form $\C \coloneqq \C - \lA \; \Dsmall \; \lAT$,
        a recursive matrix-matrix multiplication kernel that computes only a
        triangular part of $\C \coloneqq \alpha \, \lA \; \lB + \beta \, \C$.
        Note that this recursive multiplication kernel is the only update in
        \xroutine[sy]{trf} that suses a \blasl3 kernel (\xroutine{gemm}).

    \item \phantomsection\xroutine[sy]{trf\_rook}:\label{routine:xsytrf_rook}
        Alternative algorithm for the LDL decomposition using the bounded
        Bunch-Kaufman (``rook'') diagonal pivoting method.

    \item \phantomsection\xroutine{getrf}:\label{routine:xgetrf}
        LU decomposition of a general matrix with pivoting: $\lP \; \lL \lU
        \coloneqq \lA$.  Among others, \xroutine{getrf} is used with general
        matrices for inversions (\xroutine{getri}) and solutions of linear
        systems (e.g., \xroutine{gesv}).

    \item \phantomsection\xroutine{trsyl}:\label{routine:xtrsyl}
        Solution of the quasi-triangular\footnote{%
            $A$ and $B$ are in Schur canonical form and may contain $2 \times 2$
            diagonal blocks.
        } Sylvester equation $\sA \; \sX \pm \sX \; \sB = \sC$ for $\sX$.  This
        routine is used to reorder Schur factorizations and estimate the
        condition number of eigenvalue problems (\xroutine{trsen})  and arises
        on its own in systems and control theory.

        {\em Implementation Note}:
        \lapack's \xroutine{trsyl} is in itself unblocked; hence, since
        \relapack replaces \xroutine{trsyl} with a recursive algorithm requiring
        an unblocked version of the routine, a duplicate of the original routine
        is introduced under the name \xroutine{trsy2}.  Furthermore, on each
        recursion level, the input matrix $\sC$ is only split along its larger
        dimension, thus maximizing the overall size of the invoked \blasl3
        kernel \xroutine{gemm}.

    \item \phantomsection\xroutine{tgsyl}:\label{routine:xtgsyl}
        Solution of the generalized Sylvester equations $\sA \; \sR - \sL \; \sB
        = \sC$ and $\sD \; \sR - \sL \; \sE = \sF$ for $\sR$ and $\sL$.  It is
        also used to reorder Schur factorizations of matrix pairs and  in
        condition number estimations (\xroutine{tgsen}).

        {\em Implementation Note}:
        %While \xroutine{tgsyl} handles all different job types (via the {\tt
        %ijob} argument), \xroutine{tgsyl\_rec}'s interface closely resembles
        %that of \lapack's unblocked \xroutine{tgsy2}.
        Just as \xtrsyl, \xroutine{tgsyl} splits the input matrices $\sC$ and
        $\sF$ along their larger dimension, thus maximizing the size of the
        generated recursive sub-problems and calls to \xroutine{gemm}.

\end{itemize}
At compile time, \relapack allows to set the recursive-to-unblocked crossover
size~$c$ either globally or individually for each routine (default: $c = 24$).
Furthermore, each routine can be separately excluded from the generated library
{\tt librelapack.a} to allow for any mixing of \relapack and other \lapack
implementations.

\parsum{recursion not applicable: extra flops/QR-like}
To benefit from the nearly optimal efficiency of \blasl3 kernels, several
operations are implemented in \lapack as blocked algorithms at the cost of
$O(n^2 b)$ extra FLOPs with respect to the unblocked algorithm.  Translated to a
recursive algorithm, this overhead increases to $O(n^3)$ extra FLOPs, hence
making recursion infeasible for large $n$ \cite{qrrec}.  The following routines
are affected.
\begin{itemize}
    \item \xroutine{geqrf}: QR decomposition\footnote{%
            The same applies to the related decompositions RQ
            (\xroutine{gerqf}), QL (\xroutine{geqlf}), LQ
            (\xroutine{gelqf}), and RZ (\xroutine{tzrzf}).
        }: $\qQ \; \qR \coloneqq \qA$.
        %Among others, these routines are involved in the solution of over- or
        %under-determined linear systems (\xroutine{gels}) and various
        %eigensolvers (e.g., \xroutine{ggev}).

    \item \xroutine[or]{mqr}:\footnote{%
            $\texttt{\it xor} \in \{\texttt{sor}, \texttt{dor}, \texttt{cun},
            \texttt{zun}\}$, i.e., the complex routines are {\tt cunmrq} and
            {\tt zunmrq}.
        } Multiplication with an orthogonal matrix as returned by the QR
        decomposition;\footnote{%
            Applies respectively to the related decompositions: RQ
            (\xroutine[or]{mrq}), QL (\xroutine[or]{mql}), LQ
            (\xroutine[or]{mlq}), and RZ (\xroutine[or]{mrz}).
        } example: $\sC \coloneqq \rQ \; \sC$.

    \item \xroutine[or]{gqr}: Construction of the full orthogonal matrix $\qQ$
        from the format returned by the QR decomposition.\footnote{%
            Applies respectively to the related decompositions: RQ
            (\xroutine[or]{grq}), QL (\xroutine[or]{gql}), and LQ
            (\xroutine[or]{glq}).
        }

    \item \xroutine[sy]{trd}: Reduction of a symmetric (Hermitian) matrix to
        tridiagonal form: 
        
        $\ \ \rQ \; \rT \; \rQT \coloneqq \A$.
        %\xroutine[sy]{trd} serves as a key step in the solution of symmetric
        %(Hermitian) eigenvalue problems (e.g. \xroutine[sy]{ev}).

    \item \xroutine{gehrd}: Reduction of a matrix to upper Hessenberg form: $\rQ
        \; \rH \; \rQT \coloneqq \A$.
        %This reduction is used to solve general eigenvalue problems (e.g.
        %\xroutine{geev}).

    \item \xroutine{gebrd}: Reduction of a matrix to bidiagonal form: $\rQ \;
        \rB \; \rPT \coloneqq \A$.
        % This routine is used in the computation of singular value decompositions
        % (e.g., \xroutine{gesdd}) and the minimization of least squares problems
        % (e.g., \xroutine{gelsd}).
\end{itemize}

\parsum{recursion not applicable: banded}
Furthermore, recursion cannot be applied efficiently to operations on banded
matrices, since the bandwidths limit the traversal block size.  This applies to
\begin{itemize}
    \item \xroutine{pbtrf}: Banded Cholesky decomposition; example: $\bL \bLT
        \coloneqq \bA$, and
    \item \xroutine{gbtrf}: Banded LU decomposition: $\bL \bU \coloneqq
        \bAdiff$.
\end{itemize}

\parsum{missing: pivoting Chol}
To the best of our knowledge, only one \lapack operation to which recursion
might be applicable is yet not covered in \relapack:
\begin{itemize}
    \item \xroutine{pstrf}: Cholesky decomposition of a symmetric (Hermetian)
        semi-definite matrix with complete pivoting: $\lP \; \L \LT \lPT
        \coloneqq \A$.

        \xroutine{pstrf} is uses a pivoting representation different from those
        found in other routines such as \xpotrf and is thus incompatiabl with
        other \lapack operations and in fact not used anywhere throught the
        library.
\end{itemize}

\parsum{conclusion: almost perfect coverage}
To summarize, with the exception of \xroutine{pstrf}, \relapack covers all of
\lapack's compute routines to which recursion is applicable and promising to
yield performance benefits.  Due to \lapack's layered design, the performance
benefits of these recursive routines extend to many other operations.
}

%%% Local Variables:
%%% mode: latex
%%% TeX-master: "main"
%%% End:

    \section{ReLAPACK vs. Optimized Libraries}
    \label{sec:vslibs}
    \begin{figure}[p]
    \centering\small

    \pgfplotsset{
        colormap={speedup}{
            color(0.0cm)=(plot1);
            color(0.5cm)=(graybg);
            color(1.5cm)=(plot2); 
        },
        every axis/.append style={
            scale only axis,
            height=1.75cm,
            width=5.5cm,
            view={0}{90},
            xlabel={},
            xticklabels={},
            ymin={},
            ytick={0,...,32},
            yticklabels={},
            yticklabel style={font=\tt, anchor=north east},
            y dir=reverse,
            point meta min=.5,
            point meta max=2,
            anchor=north,
            yshift=-\smallskipamount,
            at=(tmp.south),
            name=tmp,
            ylabel style={rotate=-90, font=\small},
        },
    }

    \tikzset{external/export=true}
    \noindent\hphantom{(a) \xlauum}speedup:
    \tikz[baseline]{
        \node (tmp) {};
        \node[inner sep=0]{\tikzset{inner sep=.3333em}\ref*{fig:vslibfull:colorbar}};
    }
    \hphantom{speedup:$00$}

    \tikzset{external/export=true}\tikzsetnextfilename{vslibfull}
    \begin{tikzpicture}
        \node (tmp) at (0, 0) {1 thread};
        \foreach \ylabel/\y/\yticklabels/\xlabel/\xticklabel in {
            a/0/{slauum, dlauum, clauum, zlauum}//{\strut},
            b/4/{ssygst, dsygst, chegst, zhegst}//{\strut},
            c/8/{strtri, dtrtri, ctrtri, ztrtri}//{\strut},
            d/12/{spotrf, dpotrf, cpotrf, zpotrf}//{\strut},
            e/16/{ssytrf, dsytrf, chetrf, zhetrf}//{\strut},
            f/20/{sgetrf, dgetrf, cgetrf, zgetrf}//{\strut},
            g/24/{strsyl, dtrsyl, ctrsyl, ztrsyl}//{\strut},
            h/28/{stgsyl, dtgsyl, ctgsyl, ztgsyl}/{matrix size $n$}/{}
        } {
            \pgfmathparse{\y+3}\let\ya\pgfmathresult
            \pgfmathparse{\y+4}\let\yb\pgfmathresult
            \begin{axis}[
                restrict y to domain=\y:\yb,
                ytick={\y,...,\ya},
                ylabel/.expanded={(\ylabel)},
                yticklabels/.expanded=\yticklabels,
                xticklabel/.expand once=\xticklabel,
                xlabel/.expand once=\xlabel,
                colorbar horizontal,
                colorbar style={
                    xtick={.5, 1, 1.5, 2},
                    xticklabels={
                        $\leq 0.5 \phantom{\leq}$, 
                        $1$, 
                        $1.5$, 
                        $\phantom{\leq} 2 \leq$
                    }
                },
                colorbar to name=fig:vslibfull:colorbar,
            ]
                \addplot3[surf, shader=flat, mesh/interior colormap name=speedup] 
                    file {figures/data/vslibs/Haswell_MKL_lsf.1.3dat};
            \end{axis}
        }

        \node (tmp) at (6, 0) {12 threads};
        \foreach \y/\xlabel/\xticklabel in {
            0//{\strut},
            4//{\strut},
            8//{\strut},
            12//{\strut},
            16//{\strut},
            20//{\strut},
            24//{\strut},
            28/{matrix size $n$}/{}
        } {
            \pgfmathparse{\y+4}\let\yb\pgfmathresult
            \begin{axis}[
                restrict y to domain=\y:\yb,
                xticklabel/.expand once=\xticklabel,
                xlabel/.expand once=\xlabel,
            ]
                \addplot3[surf, shader=flat, mesh/interior colormap name=speedup] 
                    file {figures/data/vslibs/Haswell_MKL_lsf.12.3dat};
            \end{axis}
        }
    \end{tikzpicture}

    \caption{Speedup of \relapack over MKL on a {\sc Haswell} processor.}
    \label{fig:vslibsfull}
\end{figure}
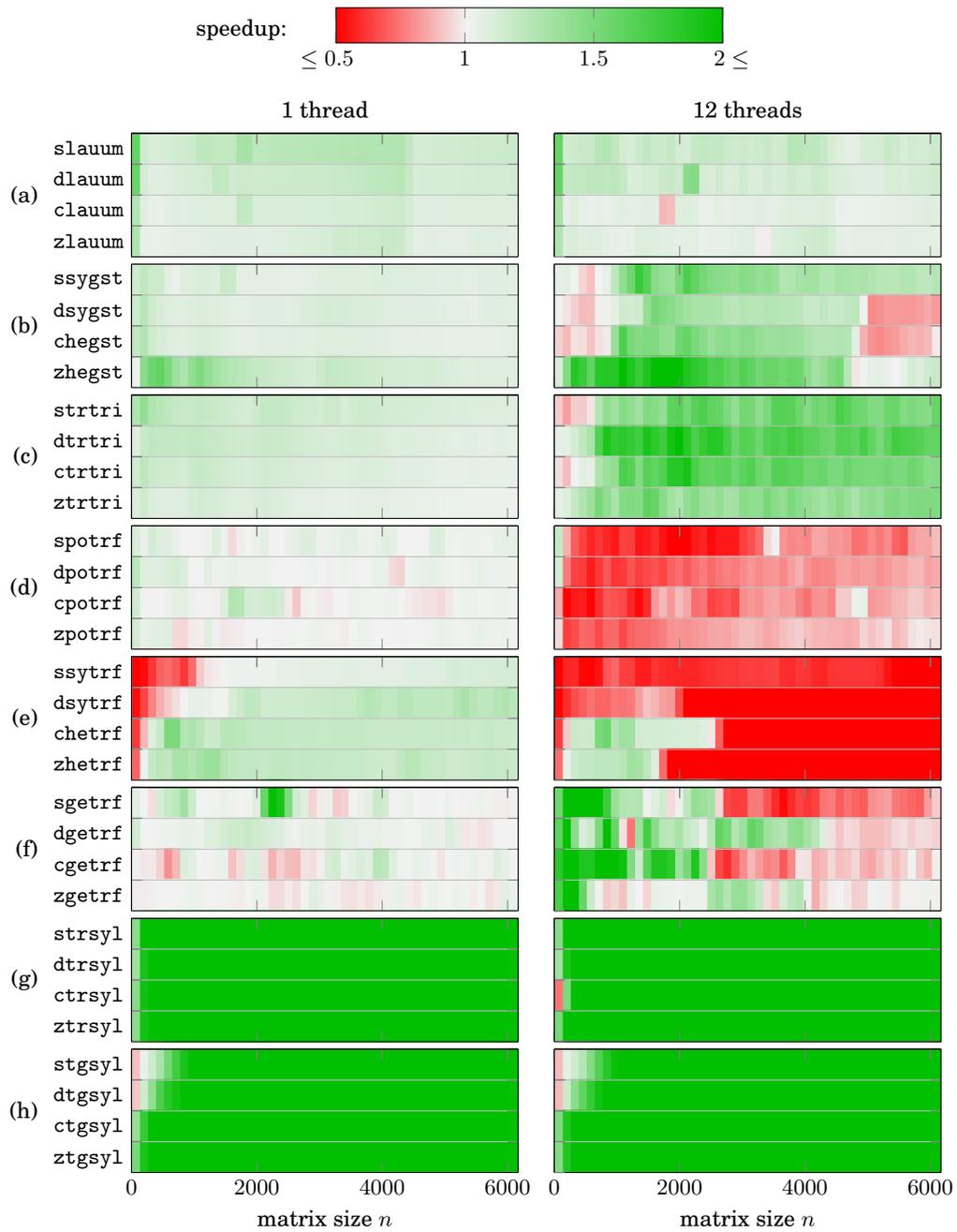

%%% Local Variables:
%%% mode: latex
%%% TeX-master: "../main"
%%% End:

\parsum{Haswell MKL setup}
While in \autoref{sec:bvsr}, we compared recrusive algorithms with blocked
algorithms using the unblocked kernels from the reference \lapack
implementation, we conclude this study with a performance comparison between
\relapack and two optimized \lapack implementations.  We start with a comparison
of all of \relapack's operations\footnote{%
    For all operations, we use the following flag arguments (where applicable):
    $\mathtt{uplo} = \mathtt L$, $\mathtt{trans} = \mathtt{tranA} =
    \mathtt{tranB} = \mathtt{diag} = \mathtt N$, and $\mathtt{itype} =
    \mathtt{ijob} = \mathtt{isgn} = 1$.
} in all four data types against {\sc Intel MKL}.  (Note that \relapack's
routines are also linked to MKL, and therefore may also benefit from
optimizations of the unblocked \lapack kernels.) We present the speedup of the
\relapack routines over the corresponding MKL routines computed as follows:
$$
    \text{speedup} = \frac{\text{time(MKL)}}{\text{time(\relapack)}} \enspace.
$$ 
\autoref{fig:vslibsfull} presents such speedups on an {\sc Intel Haswell-EP
E5-2680~v3} using only 1~core (left) and all of its 12~cores (right).

\parsum{Sylvester}
First off, we notice that for all data types and both in the single- and
multi-threaded scenario, \relapack's Sylvester solvers \xtrsyl and \xtgsyl
(\autoref{fig:vslibsfull}g and h) clearly outperforms MKL's routines: On
average, the speedup for \xtrsyl on 1 and 12~cores is, respectively, $40$ and
$100$; for \xtgsyl, it is around $9$ on 1~core, while on 12~cores, it ranges
from from $6.7$ (for double real) to $17$ (for double complex).  The numbers
suggest that in MKL these routines are not optimized are substantially
equivalent to the unblocked \lapack reference implementation. 

\parsum{single threaded}
Focusing on the single-threaded case, with only very few exceptions, the average
speedup is \num{1.0940}, i.e., using \relapack on top of MKL pays off, yielding
a performance improvement of \SI{9.40}{\percent}; it appears that only the LDL
decomposition \xsytrf (\autoref{fig:vslibsfull}e) for small matrices is
slower than MKL.

\parsum{12-threaded}
With 12~threads the scenario is different; across the board the speedups
are less uniform, ranging from \num{.5} to above \num{1.5}, and are considerably
prone to fluctuations.  For \xlauum, \xsygst (with the exception of large
matrices for double real and single complex), and \xtrtri
(\autoref{fig:vslibsfull}a, b, and c), \relapack is overall faster than MKL,
averaging a speedup of \SI{30.53}{\percent}.  For \xpotrf and, with large
matrices (beyond $n = 2500$ in single precision and $n = 4250$ in double
precision), \xgetrf (\autoref{fig:vslibsfull}d and f), \relapack is clearly
slower than MKL, since the latter employs alternative algorithmic schemes such
as algorithms-by-blocks; these schemes are specially tailored for multi-core
architectures and {\sc Intel} has put considerable effort into optimizing them.
Finally, the main cause for \relapack's poor multi-threaded performance for
\xsytrf (\autoref{fig:vslibsfull}e) is its recursive matrix multiplication
routine (see \hyperref[routine:xsytrf]{\autoref*{sec:relapack}}, \xsytrf), which
cannot match the parallelism of an optimized version close to \xroutine{gemm}
likely used by MKL.

\parsum{conclusion}
To summarize, with the exception of \xpotrf, \xsytrf, and (for large matrices)
\xgetrf, which are easily excluded at compile time, even when working with a
highly-optimized library such as MKL, it pays off to employ \relapack.

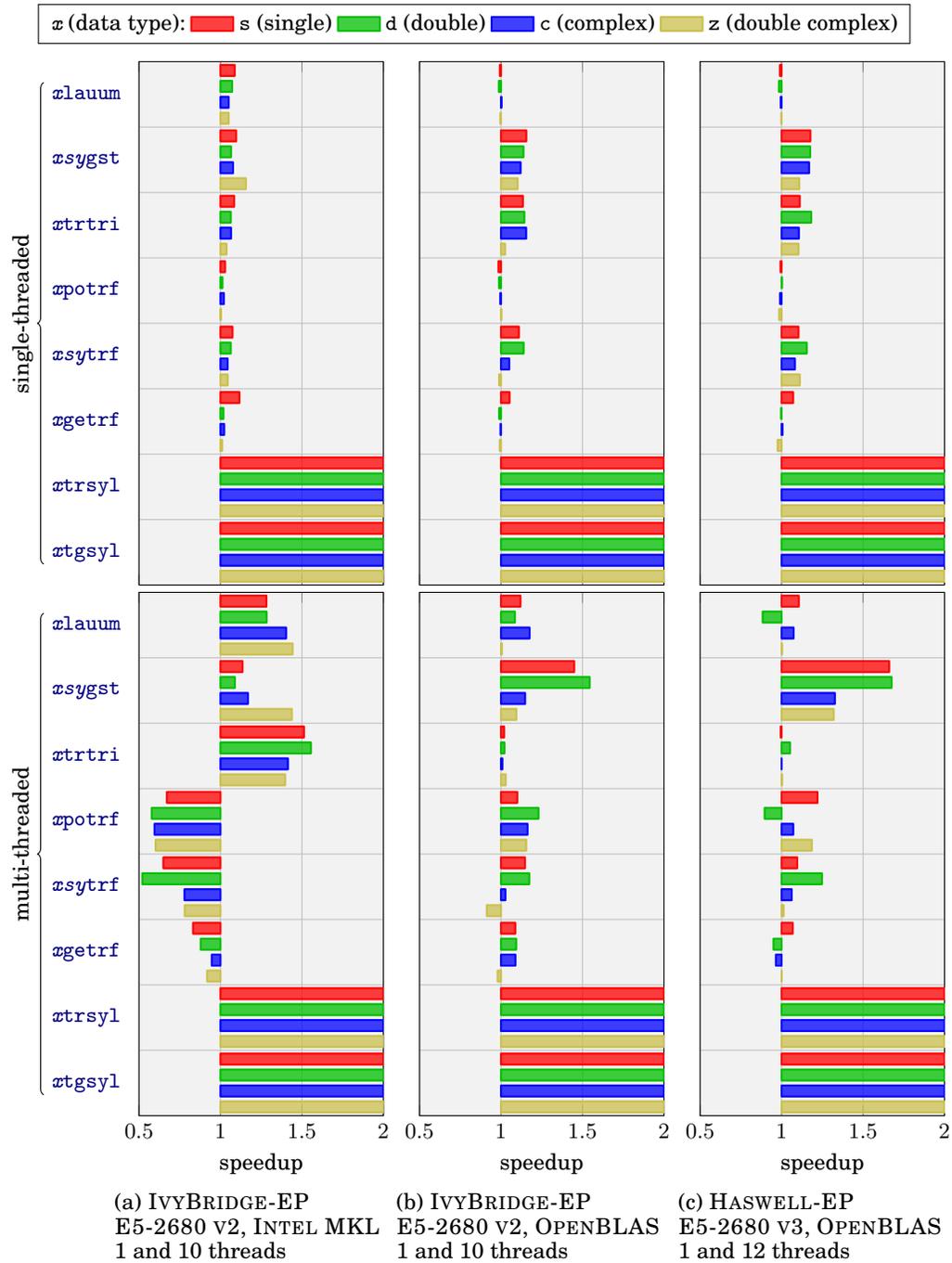
\begin{figure}[p]
    \centering\small

    \ref*{leg:vslibavg}

    \vspace\medskipamount

    \pgfplotsset{
        every axis/.append style={
            scale only axis,
            height=7.5cm,
            width=3.5cm,
            xmin=-.5,
            xmax=1,
            xlabel={speedup},
            xticklabel={\pgfmathparse{\tick+1}$\pgfmathprintnumber{\pgfmathresult}$},
            restrict x to domain*=-inf:1,
            xmajorgrids=true,
            ytick=data,
            yticklabels={},
            y tick style={draw=none},
            minor y tick num=1,
            ymajorgrids=false,
            yminorgrids=true,
            ymin=.5,
            ymax=8.5,
            ylabel shift=-1cm,
            y dir=reverse,
            xbar,
            bar width=4.5pt,
            cycle list={plot4, plot3, plot2, plot1},
        },
        every axis plot/.append style={fill, fill opacity=.75},
        first/.style={name=first, xticklabels={}},
        second/.style={anchor=north, yshift=-\smallskipamount, at=(first.south)}
    }

    \begin{subfigure}{.42\textwidth}
        \raggedleft
        \pgfplotsset{
            every axis/.append style={yticklabels={
                    \strut\xlauum, \xsygst, \strut\xtrtri, \xpotrf,
                    \xsytrf, \xgetrf, \xtrsyl, \xtgsyl,
            }},
        }
        \tikzset{external/export=true}\tikzsetnextfilename{vslibavg_IvyBridge_MKL}
        \begin{tikzpicture}
            \begin{axis}[first,
                area legend,
                reverse legend,
                legend to name=leg:vslibavg,
                legend columns=-1,
            ]
                \legend{
                    {\tt z} (double complex),
                    {\tt c} (complex),
                    {\tt d} (double),
                    {\tt s} (single),
                    {\tt\it x} (data type):
                }
                \foreach \dt in {z, c, d, s}
                    \addplot table[y expr=\lineno, x expr=\thisrow{\dt}-1]
                        {figures/data/vslibs/IvyBridge_MKL_lsf.1.avg};
                \addlegendimage{empty legend} 
                \coordinate (t) at (yticklabel cs:.96);
                \coordinate (b) at (yticklabel cs:.04);
                % xtrsyl 1
                \node[above left, plot1] at (-.25,  7) {\small\num{20}};
                \node[above left, plot2] at (0,     7) {\small\num{15}};
                \node[below left, plot3] at (-.25,  7) {\small\num{13}};
                \node[below left, plot4] at (0,     7) {\small\num{8}};
                % xtgsyl 1
                \node[above left, plot1] at (-.25,  8) {\small\num{7}};
                \node[above left, plot2] at (0,     8) {\small\num{6}};
                \node[below left, plot3] at (-.25,  8) {\small\num{14}};
                \node[below left, plot4] at (0,     8) {\small\num{10}};
            \end{axis}
            \draw[decorate, decoration={brace}] (b) -- (t) 
                node[midway, above=4pt, rotate=90] {single-threaded\strut};
            \begin{axis}[second]
                \foreach \dt in {z, c, d, s}
                    \addplot table[y expr=\lineno, x expr=\thisrow{\dt}-1]
                        {figures/data/vslibs/IvyBridge_MKL_lsf.10.avg};
                \coordinate (t) at (yticklabel cs:.96);
                \coordinate (b) at (yticklabel cs:.04);
                % xtrsyl 10
                \node[above left, plot1] at (-.25, 7) {\small\num{34}};
                \node[above left, plot2] at (0,    7) {\small\num{32}};
                \node[below left, plot3] at (-.25, 7) {\small\num{38}};
                \node[below left, plot4] at (0,    7) {\small\num{26}};
                % xtgsyl 10
                \node[above left, plot1] at (-.25, 8) {\small\num{8}};
                \node[above left, plot2] at (0,    8) {\small\num{6}};
                \node[below left, plot3] at (-.25, 8) {\small\num{25}};
                \node[below left, plot4] at (0,    8) {\small\num{24}};
            \end{axis}
            \draw[decorate, decoration={brace}] (b) -- (t) 
                node[midway, above=4pt, rotate=90] {multi-threaded\strut};
        \end{tikzpicture}

        \hfill
        \begin{minipage}{.6905\textwidth}
            \caption{\raggedright
                {\sc IvyBridge-EP E5-2680~v2}, {\sc Intel MKL}
                1~and~10~threads
            }
            \label{fig:vslibavg_IvyBridge_MKL}
        \end{minipage}
    \end{subfigure}%
    \pgfplotsset{
        every axis/.append style={yticklabel style={inner sep=0, outer sep=0}}
    }%
    \begin{subfigure}{.29\textwidth}
        \centering
        \tikzset{external/export=true}\tikzsetnextfilename{vslibavg_IvyBridge_OpenBLAS}
        \begin{tikzpicture}
            \begin{axis}[first]
                \foreach \dt in {z, c, d, s}
                    \addplot table[y expr=\lineno, x expr=\thisrow{\dt}-1]
                        {figures/data/vslibs/IvyBridge_OpenBLAS_lsf.1.avg};
                % xtrsyl 1
                \node[above left, plot1] at (-.25,  7) {\small\num{18}};
                \node[above left, plot2] at (0,     7) {\small\num{16}};
                \node[below left, plot3] at (-.25,  7) {\small\num{13}};
                \node[below left, plot4] at (0,     7) {\small\num{5}};
                % xtgsyl 1
                \node[above left, plot1] at (-.25,  8) {\small\num{10}};
                \node[above left, plot2] at (0,     8) {\small\num{7}};
                \node[below left, plot3] at (-.25,  8) {\small\num{10}};
                \node[below left, plot4] at (0,     8) {\small\num{5}};
            \end{axis}
            \begin{axis}[second]
                \foreach \dt in {z, c, d, s}
                    \addplot table[y expr=\lineno, x expr=\thisrow{\dt}-1]
                        {figures/data/vslibs/IvyBridge_OpenBLAS_lsf.10.avg};
                % xtrsyl 10
                \node[above left, plot1] at (-.25, 7) {\small\num{5}};
                \node[above left, plot2] at (0,    7) {\small\num{6}};
                \node[below left, plot3] at (-.25, 7) {\small\num{32}};
                \node[below left, plot4] at (0,    7) {\small\num{21}};
                % xtgsyl 10
                \node[above left, plot1] at (-.25, 8) {\small\num{2}};
                \node[above left, plot2] at (0,    8) {\small\num{2}};
                \node[below left, plot3] at (-.25, 8) {\small\num{10}};
                \node[below left, plot4] at (0,    8) {\small\num{9}};
            \end{axis}
        \end{tikzpicture}

        \begin{minipage}{\textwidth}
            \caption{\raggedright
                {\sc IvyBridge-EP E5-2680~v2}, {\sc OpenBLAS}
                1~and~10~threads
            }
            \label{fig:vslibavg_IvyBridge_OpenBLAS}
        \end{minipage}
    \end{subfigure}%
    \begin{subfigure}{.29\textwidth}
        \tikzset{external/export=true}\tikzsetnextfilename{vslibavg_Haswell_OpenBLAS}
        \begin{tikzpicture}
            \begin{axis}[first]
                \foreach \dt in {z, c, d, s}
                    \addplot table[y expr=\lineno, x expr=\thisrow{\dt}-1]
                        {figures/data/vslibs/Haswell_OpenBLAS_lsf.1.avg};
                % xtrsyl 1
                \node[above left, plot1] at (-.25,  7) {\small\num{28}};
                \node[above left, plot2] at (0,     7) {\small\num{23}};
                \node[below left, plot3] at (-.25,  7) {\small\num{21}};
                \node[below left, plot4] at (0,     7) {\small\num{15}};
                % xtgsyl 1
                \node[above left, plot1] at (-.25,  8) {\small\num{11}};
                \node[above left, plot2] at (0,     8) {\small\num{9}};
                \node[below left, plot3] at (-.25,  8) {\small\num{12}};
                \node[below left, plot4] at (0,     8) {\small\num{9}};
            \end{axis}
            \begin{axis}[second]
                \foreach \dt in {z, c, d, s}
                    \addplot table[y expr=\lineno, x expr=\thisrow{\dt}-1]
                        {figures/data/vslibs/Haswell_OpenBLAS_lsf.12.avg};
                % xtrsyl 10
                \node[above left, plot1] at (-.25, 7) {\small\num{5}};
                \node[above left, plot2] at (0,    7) {\small\num{6}};
                \node[below left, plot3] at (-.25, 7) {\small\num{39}};
                \node[below left, plot4] at (0,    7) {\small\num{35}};
                % xtgsyl 10
                \node[above left, plot1] at (-.25, 8) {\small\num{2}};
                \node[above left, plot2] at (0,    8) {\small\num{2}};
                \node[below left, plot3] at (-.25, 8) {\small\num{9}};
                \node[below left, plot4] at (0,    8) {\small\num{10}};
            \end{axis}
        \end{tikzpicture}%

        \begin{minipage}{\textwidth}
            \caption{\raggedright
                {\sc Haswell-EP E5-2680~v3}, {\sc OpenBLAS}
                1~and~12~threads
            }
            \label{fig:vslibavg_Haswell_OpenBLAS}
        \end{minipage}
    \end{subfigure}

    \caption{
        \relapack speedup averaged over matrices of size up to \num{6000}.\\
        Annotations for \xtrsyl and \xtgsyl: rounded speedup $\geq 2$
    }
    \label{fig:vslibsavg}
\end{figure}

%%% Local Variables:
%%% mode: latex
%%% TeX-master: "../main"
%%% End:

\parsum{other hw/sw setups}
This conclusion is reinforced by further experiments on different
hardware/software setups: \autoref{fig:vslibsavg} presents speedups for {\sc
OpenBLAS} and MKL on both the {\sc Haswell} and {\sc IvyBridge} processors.  To
condense the information in this analysis, the speedup was sampled between $n =
24$ and $6168$ in steps of $128$, and then averaged.  While for MKL we observe
the exact same behavior on the {\sc IvyBridge}
(\autoref{fig:vslibavg_IvyBridge_MKL}) as previously on the {\sc Haswell}
(\autoref{fig:vslibsfull}), the speedups for {\sc OpenBLAS} are slightly
different, yet still predominantly larger than $1$: while for the Sylvester
solvers the average speedup is \num{11.88}, for the other routines it is
\SI{8.89}{\percent}.

%%% Local Variables:
%%% mode: latex
%%% TeX-master: "main"
%%% End:

    \section{Conclusions}
    \label{sec:conclusion}
    In this paper, we studied the performance and tuning options of both blocked and
recursive algorithms for dense linear algebra operations; we showed that blocked
algorithms require careful and expensive tuning to reach optimal performance,
while recursive algorithms attain equivalent or better performance with
virtually no tuning effort.  Motivated by this observation, and in light of the
surprising lack of a library providing such recursive algorithms, we developed
\relapack. This library offers a collection of recursive algorithms for many of
\lapack's compute kernels.  Since it preserves \lapack's established interfaces,
it integrates effortlessly into existing \lapack-based application codes.
\relapack's routines were shown not only to outperform \lapack but also to
improve upon the performance of tuned implementations from OpenBLAS and MKL.

%%% Local Variables:
%%% mode: latex
%%% TeX-master: "main"
%%% End:

    \section*{Software}
    \relapack is open source (MIT license) and available on GitHub:\\
    \url{http://github.com/HPAC/ReLAPACK}

    \begin{acks}
        Financial support from the Deutsche Forschungsgemeinschaft (DFG) through
        grant GSC 111 and from the Deutsche Telekom Stiftung are gratefully
        acknowledged.  Furthermore, we thank our colleagues in the HPAC research
        group for many fruitful discussions and valuable feedback.
    \end{acks}

    \bibliographystyle{ACM-Reference-Format-Journals}
    \bibliography{referenceS}


\begin{thebibliography}{00}

%%% the Web address.

\ifx \showCODEN    \undefined \def \showCODEN     #1{\unskip}     \fi
\ifx \showDOI      \undefined \def \showDOI       #1{{\tt DOI:}\penalty0{#1}\ }
  \fi
\ifx \showISBNx    \undefined \def \showISBNx     #1{\unskip}     \fi
\ifx \showISBNxiii \undefined \def \showISBNxiii  #1{\unskip}     \fi
\ifx \showISSN     \undefined \def \showISSN      #1{\unskip}     \fi
\ifx \showLCCN     \undefined \def \showLCCN      #1{\unskip}     \fi
\ifx \shownote     \undefined \def \shownote      #1{#1}          \fi
\ifx \showarticletitle \undefined \def \showarticletitle #1{#1}   \fi
\ifx \showURL      \undefined \def \showURL       #1{#1}          \fi

\bibitem[\protect\citeauthoryear{Andersen, Gustavson, Karaivanov, Marinova,
  Wa{\'{s}}niewski, and Yalamov}{Andersen et~al\mbox{.}}{2001a}]%
        {lawra}
{Bjarne~S. Andersen}, {Fred Gustavson}, {Alexander Karaivanov}, {Minka
  Marinova}, {Jerzy Wa{\'{s}}niewski}, {and} {Plamen Yalamov}. 2001a.
\newblock {\em Applied Parallel Computing. New Paradigms for HPC in Industry
  and Academia: 5th International Workshop, PARA 2000 Bergen, Norway, June
  18--20, 2000 Proceedings}.
\newblock Springer Berlin Heidelberg, Berlin, Heidelberg, Chapter LAWRA Linear
  Algebra with Recursive Algorithms, 38--51.
\newblock
\showISBNx{978-3-540-70734-9}
\showDOI{%
\url{http://dx.doi.org/10.1007/3-540-70734-4_7}}


\bibitem[\protect\citeauthoryear{Andersen, Wa\'{s}niewski, and
  Gustavson}{Andersen et~al\mbox{.}}{2001b}]%
        {cholrecstorage}
{Bjarne~Stig Andersen}, {Jerzy Wa\'{s}niewski}, {and} {Fred~G. Gustavson}.
  2001b.
\newblock \showarticletitle{A Recursive Formulation of Cholesky Factorization
  of a Matrix in Packed Storage}.
\newblock {\em ACM Trans. Math. Softw.\/} {27}, 2 (June 2001), 214--244.
\newblock
\showISSN{0098-3500}
\showDOI{%
\url{http://dx.doi.org/10.1145/383738.383741}}


\bibitem[\protect\citeauthoryear{Anderson, Bai, Bischof, Blackford, Demmel,
  Dongarra, Du~Croz, Greenbaum, Hammarling, McKenney, and Sorensen}{Anderson
  et~al\mbox{.}}{1999}]%
        {lapack}
{E. Anderson}, {Z. Bai}, {C. Bischof}, {L. Blackford}, {J. Demmel}, {J.
  Dongarra}, {J. Du~Croz}, {A. Greenbaum}, {S. Hammarling}, {A. McKenney},
  {and} {D. Sorensen}. 1999.
\newblock {\em LAPACK Users' Guide\/} (third ed.).
\newblock Society for Industrial and Applied Mathematics.
\newblock
\showDOI{%
\url{http://dx.doi.org/10.1137/1.9780898719604}}


\bibitem[\protect\citeauthoryear{Anderson and Dongarra}{Anderson and
  Dongarra}{1990}]%
        {lapackblocked}
{Ed Anderson} {and} {Jack Dongarra}. 1990.
\newblock {\em Evaluating Block Algorithm Variants in LAPACK}.
\newblock {T}echnical {R}eport. Philadelphia, PA, USA. 3--8 pages.
\newblock
\showISBNx{0-89871-262-9}
\showURL{%
\url{http://dl.acm.org/citation.cfm?id=645819.669692}}


\bibitem[\protect\citeauthoryear{Badia, Herrero, Labarta, Pérez,
  Quintana-Ortí, and Quintana-Ortí}{Badia et~al\mbox{.}}{2009}]%
        {smpssdla}
{Rosa~M. Badia}, {José~R. Herrero}, {Jesús Labarta}, {Josep~M. Pérez},
  {Enrique~S. Quintana-Ortí}, {and} {Gregorio Quintana-Ortí}. 2009.
\newblock \showarticletitle{Parallelizing dense and banded linear algebra
  libraries using SMPSs}.
\newblock {\em Concurrency and Computation: Practice and Experience\/} {21}, 18
  (2009), 2438--2456.
\newblock
\showISSN{1532-0634}
\showDOI{%
\url{http://dx.doi.org/10.1002/cpe.1463}}


\bibitem[\protect\citeauthoryear{Bientinesi, Gunter, and van~de
  Geijn}{Bientinesi et~al\mbox{.}}{2008}]%
        {spdinv}
{Paolo Bientinesi}, {Brian Gunter}, {and} {Robert~A. van~de Geijn}. 2008.
\newblock \showarticletitle{Families of Algorithms Related to the Inversion of
  a Symmetric Positive Definite Matrix}.
\newblock {\em ACM Trans. Math. Softw.\/} {35}, 1, Article 3 (July 2008), 22
  pages.
\newblock
\showISSN{0098-3500}
\showDOI{%
\url{http://dx.doi.org/10.1145/1377603.1377606}}


\bibitem[\protect\citeauthoryear{Bosilca, Bouteiller, Danalis, Herault,
  Lemarinier, and Dongarra}{Bosilca et~al\mbox{.}}{2012}]%
        {dague}
{George Bosilca}, {Aurelien Bouteiller}, {Anthony Danalis}, {Thomas Herault},
  {Pierre Lemarinier}, {and} {Jack Dongarra}. 2012.
\newblock \showarticletitle{DAGuE: A generic distributed {DAG} engine for High
  Performance Computing}.
\newblock {\it Parallel Comput.} {38}, 1–2 (2012), 37 -- 51.
\newblock
\showISSN{0167-8191}
\showDOI{%
\url{http://dx.doi.org/10.1016/j.parco.2011.10.003}}
\newblock
\shownote{Extensions for Next-Generation Parallel Programming Models.}


\bibitem[\protect\citeauthoryear{Brodal}{Brodal}{2004}]%
        {cacheoblivious2}
{Gerth~St{\o}lting Brodal}. 2004.
\newblock {\em Algorithm Theory - SWAT 2004: 9th Scandinavian Workshop on
  Algorithm Theory, Humleb{\ae}k, Denmark, July 8-10, 2004. Proceedings}.
\newblock Springer Berlin Heidelberg, Berlin, Heidelberg, Chapter
  Cache-Oblivious Algorithms and Data Structures, 3--13.
\newblock
\showISBNx{978-3-540-27810-8}
\showDOI{%
\url{http://dx.doi.org/10.1007/978-3-540-27810-8_2}}


\bibitem[\protect\citeauthoryear{Chan, Quintana-Orti, Quintana-Orti, and van~de
  Geijn}{Chan et~al\mbox{.}}{2007}]%
        {supermatrix}
{Ernie Chan}, {Enrique~S. Quintana-Orti}, {Gregorio Quintana-Orti}, {and}
  {Robert van~de Geijn}. 2007.
\newblock \showarticletitle{Supermatrix Out-of-order Scheduling of Matrix
  Operations for SMP and Multi-core Architectures}. In {\em Proceedings of the
  Nineteenth Annual ACM Symposium on Parallel Algorithms and Architectures}
  {\em (SPAA '07)}. ACM, New York, NY, USA, 116--125.
\newblock
\showISBNx{978-1-59593-667-7}
\showDOI{%
\url{http://dx.doi.org/10.1145/1248377.1248397}}


\bibitem[\protect\citeauthoryear{Dongarra and Ostrouchov}{Dongarra and
  Ostrouchov}{1990}]%
        {factorizationtuning}
{Jack Dongarra} {and} {Susan Ostrouchov}. 1990.
\newblock {\em LAPACK Working Note 24: LAPACK Block Factorization Algorithms on
  the INtel iPSC/860}.
\newblock {T}echnical {R}eport. Knoxville, TN, USA.
\newblock


\bibitem[\protect\citeauthoryear{Dongarra, Cruz, Hammerling, and Duff}{Dongarra
  et~al\mbox{.}}{1990}]%
        {blas3}
{J.~J. Dongarra}, {Jermey~Du Cruz}, {Sven Hammerling}, {and} {I.~S. Duff}.
  1990.
\newblock \showarticletitle{Algorithm 679: A Set of Level 3 Basic Linear
  Algebra Subprograms: Model Implementation and Test Programs}.
\newblock {\em ACM Trans. Math. Softw.\/} {16}, 1 (March 1990), 18--28.
\newblock
\showISSN{0098-3500}
\showDOI{%
\url{http://dx.doi.org/10.1145/77626.77627}}


\bibitem[\protect\citeauthoryear{Du~Croz and Higham}{Du~Croz and
  Higham}{1992}]%
        {trtristability}
{Jeremy~J Du~Croz} {and} {Nicholas~J Higham}. 1992.
\newblock \showarticletitle{Stability of Methods for Matrix Inversion}.
\newblock {\it IMA J. Numer. Anal.} {12}, 1 (1992), 1--19.
\newblock
\showDOI{%
\url{http://dx.doi.org/10.1093/imanum/12.1.1}}


\bibitem[\protect\citeauthoryear{Elmroth and Gustavson}{Elmroth and
  Gustavson}{2000}]%
        {qrrec}
{E. Elmroth} {and} {F.G. Gustavson}. 2000.
\newblock \showarticletitle{Applying recursion to serial and parallel QR
  factorization leads to better performance}.
\newblock {\em IBM Journal of Research and Development\/} {44}, 4 (July 2000),
  605--624.
\newblock
\showISSN{0018-8646}
\showDOI{%
\url{http://dx.doi.org/10.1147/rd.444.0605}}


\bibitem[\protect\citeauthoryear{Frigo, Leiserson, Prokop, and
  Ramachandran}{Frigo et~al\mbox{.}}{1999}]%
        {cacheoblivious1}
{Matteo Frigo}, {Charles~E. Leiserson}, {Harald Prokop}, {and} {Sridhar
  Ramachandran}. 1999.
\newblock \showarticletitle{Cache-Oblivious Algorithms}. In {\em Proceedings of
  the 40th Annual Symposium on Foundations of Computer Science} {\em (FOCS
  '99)}. IEEE Computer Society, Washington, DC, USA, 285--.
\newblock
\showISBNx{0-7695-0409-4}
\showURL{%
\url{http://dl.acm.org/citation.cfm?id=795665.796479}}


\bibitem[\protect\citeauthoryear{Georgiev and Wa{\'{s}}niewski}{Georgiev and
  Wa{\'{s}}niewski}{2001}]%
        {lurec}
{K. Georgiev} {and} {J. Wa{\'{s}}niewski}. 2001.
\newblock {\em Numerical Analysis and Its Applications: Second
  InternationalConference, NAA 2000 Rousse, Bulgaria, June 11--15, 2000 Revised
  Papers}.
\newblock Springer Berlin Heidelberg, Berlin, Heidelberg, Chapter Recursive
  Version of LU Decomposition, 325--332.
\newblock
\showISBNx{978-3-540-45262-1}
\showDOI{%
\url{http://dx.doi.org/10.1007/3-540-45262-1_38}}


\bibitem[\protect\citeauthoryear{Gustavson}{Gustavson}{1997}]%
        {dlarec}
{F.G. Gustavson}. 1997.
\newblock \showarticletitle{Recursion leads to automatic variable blocking for
  dense linear-algebra algorithms}.
\newblock {\em IBM Journal of Research and Development\/} {41}, 6 (Nov 1997),
  737--755.
\newblock
\showISSN{0018-8646}
\showDOI{%
\url{http://dx.doi.org/10.1147/rd.416.0737}}


\bibitem[\protect\citeauthoryear{Gustavson, Henriksson, Jonsson,
  K{\aa}gstr{\"o}m, and Ling}{Gustavson et~al\mbox{.}}{1998}]%
        {recstorage}
{Fred Gustavson}, {Andr{\'e} Henriksson}, {Isak Jonsson}, {Bo
  K{\aa}gstr{\"o}m}, {and} {Per Ling}. 1998.
\newblock {\em Applied Parallel Computing Large Scale Scientific and Industrial
  Problems: 4th International Workshop, PARA'98 Ume{\aa}, Sweden, June 14--17,
  1998 Proceedings}.
\newblock Springer Berlin Heidelberg, Berlin, Heidelberg, Chapter Recursive
  blocked data formats and BLAS's for dense linear algebra algorithms,
  195--206.
\newblock
\showISBNx{978-3-540-49261-0}
\showDOI{%
\url{http://dx.doi.org/10.1007/BFb0095337}}


\bibitem[\protect\citeauthoryear{Jonsson and K{\aa}gstr{\"o}m}{Jonsson and
  K{\aa}gstr{\"o}m}{2003}]%
        {recsy}
{Isak Jonsson} {and} {Bo K{\aa}gstr{\"o}m}. 2003.
\newblock {\em Euro-Par 2003 Parallel Processing: 9th International Euro-Par
  Conference Klagenfurt, Austria, August 26-29, 2003 Proceedings}.
\newblock Springer Berlin Heidelberg, Berlin, Heidelberg, Chapter RECSY -- A
  High Performance Library for Sylvester-Type Matrix Equations, 810--819.
\newblock
\showISBNx{978-3-540-45209-6}
\showDOI{%
\url{http://dx.doi.org/10.1007/978-3-540-45209-6_111}}


\bibitem[\protect\citeauthoryear{Karlsson}{Karlsson}{2006}]%
        {trinvrec}
{Lars Karlsson}. 2006.
\newblock {\em Computing explicit matrix inverses by recursion}.
\newblock Ph.D. Dissertation. MS thesis, Umea University, Department of
  Computing Science, Sweden.
\newblock


\bibitem[\protect\citeauthoryear{Tan, Goh, March, and See}{Tan
  et~al\mbox{.}}{2009}]%
        {hpltuning}
{Tuan~Zea Tan}, {R.S.M. Goh}, {V. March}, {and} {S. See}. 2009.
\newblock \showarticletitle{Data mining analysis to validate performance tuning
  practices for HPL}. In {\em Cluster Computing and Workshops, 2009. CLUSTER
  '09. IEEE International Conference on}. 1--8.
\newblock
\showISSN{1552-5244}
\showDOI{%
\url{http://dx.doi.org/10.1109/CLUSTR.2009.5289175}}


\bibitem[\protect\citeauthoryear{{V}an {Z}ee}{{V}an {Z}ee}{2009}]%
        {flame}
{Field~G. {V}an {Z}ee}. 2009.
\newblock {\em {\tt libflame}: {T}he {C}omplete {R}eference}.
\newblock {\tt lulu.com}.
\newblock


\bibitem[\protect\citeauthoryear{Wasniewski, Andersen, and
  Gustavson}{Wasniewski et~al\mbox{.}}{1998}]%
        {cholrec}
{Jerzy Wasniewski}, {Bjarne~Stig Andersen}, {and} {Fred~G. Gustavson}. 1998.
\newblock \showarticletitle{Recursive Formulation of Cholesky Algorithm in
  Fortran 90}.
\newblock In {\em Proceedings of the 4th International Workshop on Applied
  Parallel Computing, Large Scale Scientific and Industrial Problems}.
  Springer-Verlag, London, UK, UK, 574--578.
\newblock
\showISBNx{3-540-65414-3}
\showURL{%
\url{http://dl.acm.org/citation.cfm?id=645781.666513}}


\bibitem[\protect\citeauthoryear{Whaley}{Whaley}{2008}]%
        {lapacktuning}
{R.C. Whaley}. 2008.
\newblock \showarticletitle{Empirically tuning LAPACK's blocking factor for
  increased performance}. In {\em Computer Science and Information Technology,
  2008. IMCSIT 2008. International Multiconference on}. 303--310.
\newblock
\showDOI{%
\url{http://dx.doi.org/10.1109/IMCSIT.2008.4747256}}


\bibitem[\protect\citeauthoryear{Xianyi}{Xianyi}{2015}]%
        {openblas}
{Zhang Xianyi}. 2015.
\newblock {OpenBLAS}.
\newblock \url{http://www.openblas.net/}.   (2015).
\newblock


\bibitem[\protect\citeauthoryear{YarKhan, Kurzak, and Dongarra}{YarKhan
  et~al\mbox{.}}{2011}]%
        {quark}
{Asim YarKhan}, {Jakub Kurzak}, {and} {Jack Dongarra}. 2011.
\newblock \showarticletitle{Quark users’ guide: Queueing and runtime for
  kernels}.
\newblock {\em University of Tennessee Innovative Computing Laboratory
  Technical Report ICL-UT-11-02\/} (2011).
\newblock


\end{thebibliography}
\end{document}